\documentclass[]{./Outils_latex/aa}

\usepackage[utf8]{inputenc} 

\usepackage{natbib}         
\usepackage{graphicx}
\usepackage{longtable}
\usepackage{xcolor}
\usepackage{amssymb}
\usepackage[switch]{lineno} 
\usepackage{hyperref}
\usepackage{rotating}
\usepackage{subcaption}
\usepackage{sidecap}
\usepackage{comment}

\bibpunct{(}{)}{;}{a}{}{,} 

\usepackage{soul} 

\hypersetup{
colorlinks = true,
breaklinks = true,
urlcolor = blue,
linkcolor = blue,
citecolor = blue,
}

\newcommand{\ind}[1]{_{\mathrm{#1}}}
\newcommand{\diff}{\mathrm{d}}

\newcommand\Kepler{\emph{Kepler}}

\newcommand\tess{TESS}
\newcommand\corot{CoRoT}

\newcommand\mesa{MESA}

\newcommand\python{\emph{Python}}
\newcommand\corner{\emph{Corner}}
\newcommand\emcee{\emph{Emcee}}

\newcommand\numax{\nu\ind{max}}

\newcommand\Dnu{\Delta\nu}

\newcommand\Teff{T\ind{eff}}

\newcommand\alphaovHe{\alpha_{\mathrm{ov, He}}}
\newcommand\alphaovH{\alpha_{\mathrm{ov, H}}}
\newcommand\alphaMLT{\alpha_{\mathrm{MLT}}}
\newcommand\alphaundersh{\alpha_{\mathrm{ov, env}}}
\newcommand\alphath{\alpha_{\mathrm{th}}}
\newcommand\coefdiffthline{D_{\mathrm{th}}}
\newcommand\gradad{\nabla_{\mathrm{ad}}}
\newcommand\gradrad{\nabla_{\mathrm{rad}}}

\newcommand\rationumaxAGBbclump{\log \nu\ind{max,AGBb} / \log \nu\ind{max,clump}}
\newcommand\ratioTeffAGBbclump{\log T\ind{eff,AGBb} / \log T\ind{eff,clump}}

\newcommand\Omegacrit{\Omega_{\mathrm{crit}}}

\newcommand\Pbg{P_{\mathrm{bg}}}
\newcommand\sigmabg{\sigma_{\mathrm{bg}}}
\newcommand\abg{a_{\mathrm{bg}}}
\newcommand\bbg{b_{\mathrm{bg}}}
\newcommand\cbg{c_{\mathrm{bg}}}
\newcommand\Aexp{A_{\mathrm{exp}}}

\newcommand\fbiv{f_{\mathrm{biv}}}
\newcommand\fbg{f_{\mathrm{bg}}}

\usepackage{amstext}

\begin{document}

\title{Characterising the AGB bump and its potential to constrain mixing processes in stellar interiors}
\titlerunning{The AGBb}

\author{%
 G. Dr\'eau\inst{1},
 Y. Lebreton\inst{1,2},
 B. Mosser\inst{1},
  D. Bossini\inst{3},
 J. Yu\inst{4}
}

\institute{
\inst{1} LESIA, Observatoire de Paris, PSL Research University, CNRS, Universit\'e Pierre et Marie Curie,
 Universit\'e Paris Diderot,  F-92195 Meudon, France \\
 email: \texttt{Guillaume.Dreau@obspm.fr}\\
\inst{2} Univ Rennes, CNRS, IPR (Institut de Physique de Rennes) - UMR 6251, F-35000 Rennes, France\\
\inst{3} Instituto de Astrof\'isica e Ci\^encias do Espa\c co, Universidade do Porto, CAUP, Rua das Estrelas, PT4150-762 Porto, Portugal \\
\inst{4} Max Planck Institute for Solar System Research, Justus-von-Liebig-Weg 3, D-37077 G\"{o}ttingen, Germany \\
 }


\abstract{In the 90's, theoretical studies motivated the use of the asymptotic-giant branch bump (AGBb) as a standard candle given the weak dependence between its luminosity and stellar metallicity. Because of the small size of observed asymptotic-giant branch (AGB) samples, detecting the AGBb is not an easy task. However, this is now possible thanks to the wealth of data collected by the \corot, \Kepler, and \tess\ space-borne missions.}
{ It is well-known that the AGB bump provides valuable information on the internal structure of low-mass stars, particularly on mixing processes such as core overshooting during the core He-burning phase. Here, we investigate the dependence with stellar mass and metallicity of the calibration of stellar models to observations.}
{In this context, we analysed $\sim$ 4,000 evolved giants observed by \Kepler\ and \tess, including red-giant branch (RGB) stars and AGB stars, for which asteroseismic and spectrometric data are available. By using statistical mixture models, we detected the AGBb both in frequency at maximum oscillation power $\numax$ and in effective temperature $\Teff$. Then, we used the Modules for Experiments in Stellar Astrophysics (\mesa) stellar evolution code to model AGB stars and match the AGBb occurrence with observations.}
{From observations, we could derive the AGBb location in 15 bins of mass and metallicity. We noted that the higher the mass, the later the AGBb occurs in the evolutionary track, which agrees with theoretical works. Moreover, we found a slight increase of the luminosity at the AGBb when the metallicity increases. By fitting those observations with stellar models, we noticed that low-mass stars ($M \leq 1.0 M_{\odot}$) require a small core overshooting region during the core He-burning phase. This core overshooting extent increases toward high mass, but above $M \geq 1.5 M_{\odot}$ we found that the AGBb location cannot be reproduced with a realistic  He-core overshooting alone, and instead additional mixing processes have to be invoked.}
{The observed dependence on metallicity complicates the use of the AGBb as a standard candle. Moreover, different mixing processes may occur according to the stellar mass. At low mass ($M \leq 1.5M_{\odot}$), the AGBb location can be used to constrain the He-core overshooting. At high mass ($M \geq 1.5M_{\odot}$), an additional mixing induced for instance by rotation is needed to reproduce observations.}

\maketitle

\section{Introduction}

The asymptotic-giant branch (AGB) is a key stage of stellar evolution that can be used to constrain both the stellar structure and environment. On the one hand, observations of circumstellar CO line emission and stellar light scattered by dust in circumstellar envelopes allow us to estimate the mass-loss rate on the AGB, which is crucial to understand the final stages of stellar evolution and the metal enrichment in the interstellar medium, hence the chemical enrichment of galaxies \citep[e.g.][]{1998ApJS..117..209K, 2006A&A...452..257M, 2008A&A...487..645R, 2018MNRAS.481.4984M, 2019MNRAS.484.4678M}. On the other hand, the AGB provides valuable constraints for stellar interiors with the help of stellar models \citep{2015MNRAS.453.2290B}. Current stellar models suffer from systematic uncertainties due to our limited understanding of physical processes in stellar interiors. Particularly, constraining mixing processes in advanced burning stages is demanding because it requires to implement helium semiconvection to take into account the additional helium captured by the growing He-core  \citep[e.g.][]{1971Ap&SS..10..355C, 1972ApJ...171..309R, 1973BAAS....5..314S, 2017RSOS....470192S}. Then, the use of observational constraints linked to stellar interiors is crucial to test the reliability of stellar models. With this in mind, several studies aimed at constraining stellar parameters of red giants with asteroseismic observables \citep{2011MNRAS.415.3783D, 2012A&A...538A..73B, 2015A&A...580A.141L}. 
Using the global seismic parameters, i.e. the large frequency separation $\Dnu$ and the frequency of the maximum oscillation power $\numax$, these authors could infer the mass and radius of red giants and reduce their uncertainties by a factor of more than 3 compared with those based on spectroscopic constraints only. On top of these asteroseismic observables, the use of both the mode inertias and coupling factor between the g- and p- mode cavities in red giants provides unique constraints on the mode trapping, hence on the innermost stellar structure \citep{2014ApJ...781L..29B, 2020A&A...634A..68P}. However, all the studies mentioned herebefore focus on the early stages of red giants, so that additional work needs to be done to constrain stellar structure during the helium burning stages. \\

One of the key events happening in the helium burning phase that still needs to be constrained is the AGBb. This is now possible with the recent seismic constraints obtained for high-luminosity RGB and AGB stars with $\Dnu \leq 4.0~\mu$Hz \citep{2021A&A...650A.115D}. The AGBb manifests through a luminosity drop as a star evolves on the AGB and is associated with the ignition of the He-burning shell source. The AGBb is then characterised by a local excess of stars in the luminosity distribution of stellar populations. While the AGBb has been first predicted by stellar evolutionary models \citep{1978MNRAS.184..377C}, it has then been identified in the colour-magnitude diagram of a few Galactic globular clusters \citep{1992MmSAI..63..491F}. \cite{2015MNRAS.453.2290B} have shown that the AGBb can be used to constrain the core mixing scheme during the core He-burning phase. They could reproduce both the seismic constraints and the AGBb luminosity of observed \Kepler\ red clump stars by considering core overshooting of the mixed He core with a moderate value of core overshooting ($\alphaovHe = 0.5$ where $\alphaovHe$ is the ratio of the overshooting length to the pressure scale height). Using the AGBb luminosity as a stellar model constraint allows to reduce the systematic uncertainties on the mixing processes beyond the boundary of the convective envelope, which are essential to predict stellar lifetime in the core He-burning phase \citep[e.g.,][]{1971Ap&SS..10..340C, 2007IAUS..239..235C}. \\ 

The characterisation of the luminosity bump on the RGB with seismic data has already been  achieved \citep{2018ApJ...859..156K}. By combining \Kepler\ and APOGEE data of thousands of red giants, these authors highlighted that the location of the red-giant branch bump (RGBb) is sensitive to the stellar mass and metallicity. Moreover, they showed that significant overshooting from the base of the convective envelope during the main sequence must be considered to reproduce the location of the RGBb, with an efficiency that increases with decreasing metallicity. A similar description of mixing beyond the convective envelope during He-burning phases would help to predict the third dredge-up efficiency on the thermally pulsing AGB (TP-AGB) phase \citep{2000MmSAI..71..745H, 2007A&A...469..239M, 2020MNRAS.493.4748W}. Moreover, a precise characterisation of the AGBb would confirm or disprove the potential of the AGBb to be a suitable candidate for standard candles \citep{1992MmSAI..63..485P, 1992MmSAI..63..491F}. \\

In this study, we aim at detecting and characterising the AGBb. First, we investigate its dependence with the stellar mass and metallicity by using \Kepler\ and \tess\ asteroseismic targets. Then, we use the AGBb as a calibrator for mixing processes, particularly for core overshooting during the He-burning phase. The article is organised as follows. The data set is described in Sect.~\ref{sec:data_set}. In Sect.~\ref{sec:models}, we define the macrophysics and microphysics implemented to model stellar evolution up to the AGB phase. Methods used to locate and characterise the AGBb in models and observations are presented in Sect.~\ref{sec:method}. The results are analysed in Sect.~\ref{sec:results}.
They illustrate the needs to take He-core overshooting into account in stellar models to reproduce the observed location of the AGBb. We discuss our results and explore the impact of other parameters on the AGBb location in Sect.~\ref{sec:discussion}. Eventually, Sect.~\ref{sec:conclusion} is devoted to conclusions.

\section{Data set}
\label{sec:data_set}

\begin{figure*}[htbp]
	\begin{minipage}{1.\linewidth}  
		\rotatebox{0}{\includegraphics[width=0.5\linewidth]{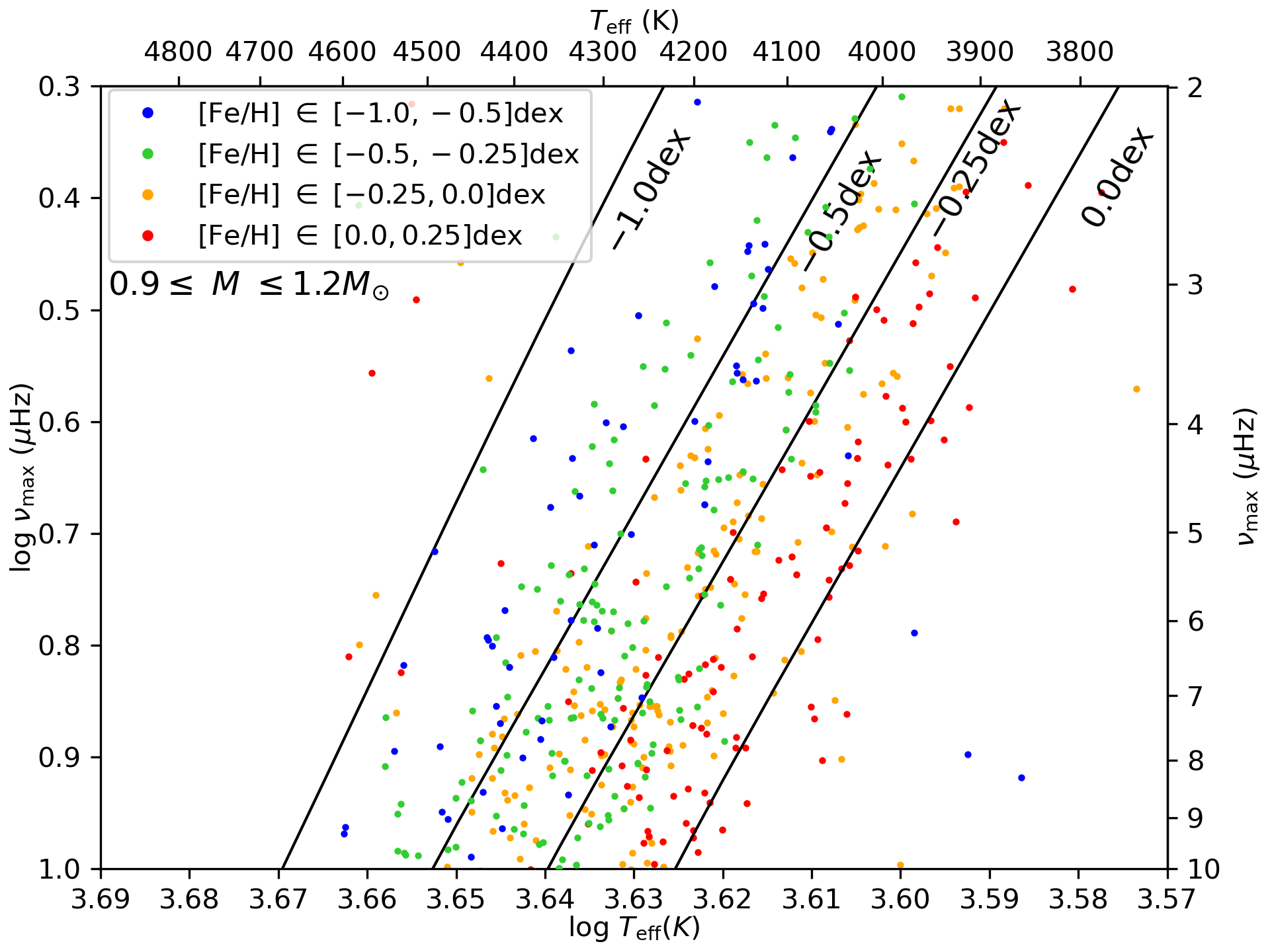}}
		\rotatebox{0}{\includegraphics[width=0.5\linewidth]{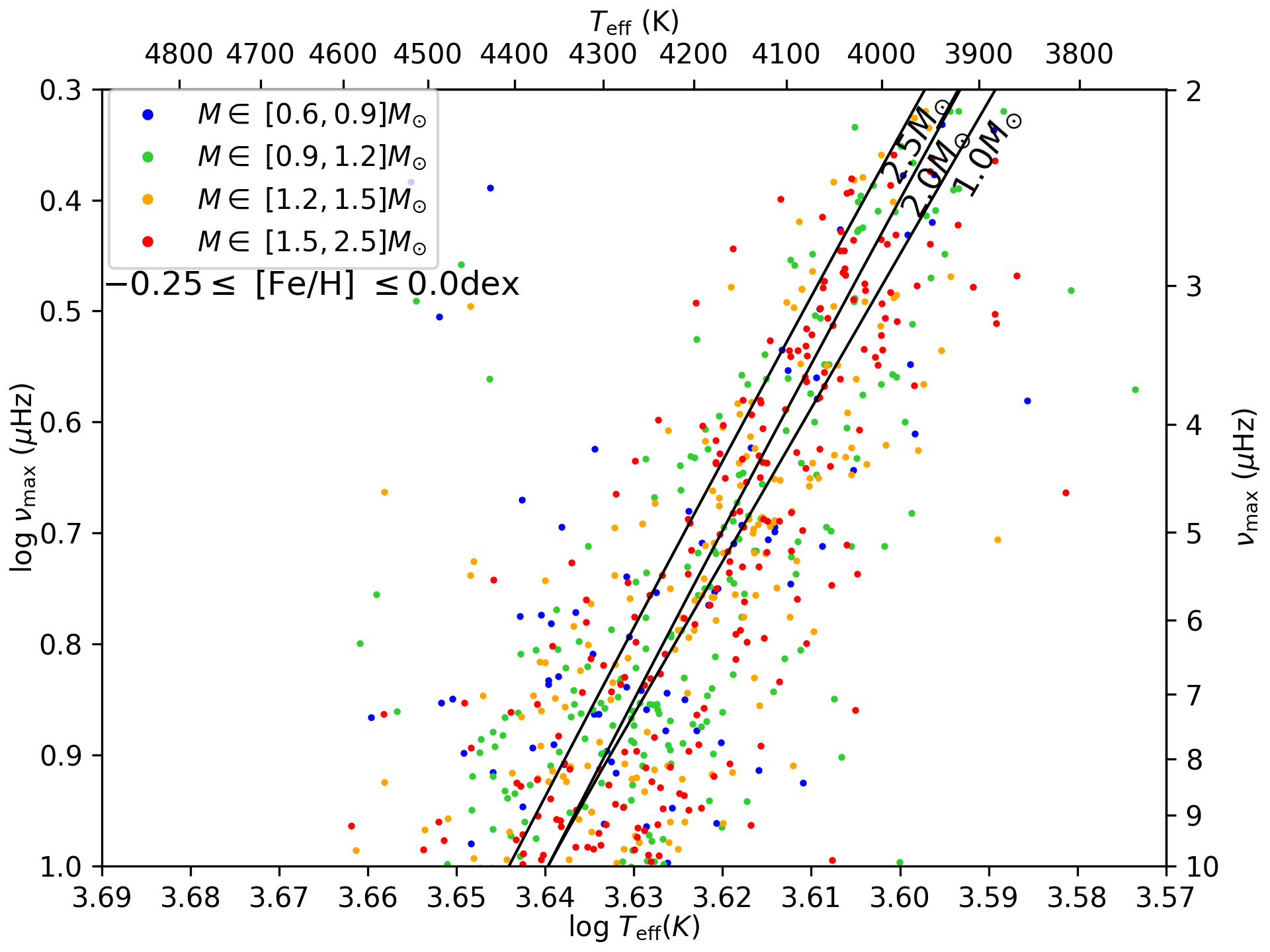}}
	\end{minipage}
	\caption{Seismic HR diagram of our sample of stars in the bin of mass $M\in [0.9, 1.2]M_{\odot}$ (left) and in the bin of metallicity [Fe/H] $\in[-0.25, 0.0]$ dex (right). In both panels, stars with different metallicities (left) and masses (right) are represented with different colours. On top of those observations, stellar evolutionary tracks along the AGB are represented with black lines at different metallicities at $1\,M_{\odot}$ (left) and masses at $\mathrm{[Fe/H]} = -0.25\,$dex (right).
	}
	\label{fig:seism_HR_diag}
\end{figure*}

In order to detect the AGBb, we selected evolved stars that have been observed by the \Kepler\ and \tess\ telescopes, including RGB and AGB stars. In order to reject red clump stars from the sample, we only kept stars with $\numax \leq 22~\mu$Hz (or equivalently $\Dnu \lesssim 2.7~\mu$Hz) because no AGBb is expected to occur above this limit \citep[][their Fig. 9]{2021A&A...650A.115D}.
It has been shown that at low $\numax \leq 10~\mu$Hz (or equivalently $\Dnu\lesssim 1.5 \mu$Hz), it is difficult to safely distinguish AGB stars from RGB ones \citep{2012A&A...541A..51K, 2019A&A...622A..76M}. On the other hand, the local excess associated to the AGBb is well visible on top of the background composed of RGB and He-burning stars \citep{2015MNRAS.453.2290B}. Therefore, since we wished to work on a sample containing as many AGB stars as possible, we decided not to reject any (suspected) RGB star from the initial sample.
Considering more stars, even unclassified, allowed us to include more AGB stars in our sample. The evolved \Kepler\ targets around this evolutionary stage have been the subject of an exhaustive seismic analysis \citep{2013A&A...559A.137M, 2014A&A...572L...5M, 2019A&A...622A..76M, 2014ApJ...788L..10S, 2020MNRAS.493.1388Y, 2021MNRAS.501.5135Y, 2021A&A...650A.115D}, providing estimates of $\numax$. Then, we used the $\numax$ estimates from \cite{2014A&A...572L...5M, 2019A&A...622A..76M} while we selected their mass $M$ from the APOKASC catalogue \citep{2014ApJS..215...19P, 2018ApJS..239...32P}. The later is a survey of \Kepler\ targets complemented by spectroscopic data. The effective temperatures $\Teff$ and stellar metallicities are taken from the catalogues of APOGEE DR17 \citep{2021arXiv211202026A}, GALAH DR3 \citep{2021MNRAS.506..150B}, and RAVE DR6 \citep{2020AJ....160...83S}.
We took the stellar masses derived from the semi-empirical asteroseismic scaling relation presented in \cite{1995A&A...293...87K}, and corrected by a factor that is adjusted star by star when available\footnote{This correction is applied to capture the deviations from the asteroseismic scaling relations between the stellar mass $M$, radius $R$, large frequency separation $\Dnu$, and frequency at maximum oscillation power $\numax$.} \citep{2018ApJS..239...32P}. When stellar masses are not available, which concerns about 10\% of our \Kepler\ targets, we estimated them with the semi-empirical relation without any correction factor. As for the \tess\ targets, they have been studied in \cite{2021MNRAS.502.1947M}. We picked $\numax$ as the mean value between three pipelines \citep{2009A&A...508..877M, 2010A&A...511A..46M, 2020RNAAS...4..177E}. We extracted the mass $M$, effective temperature $\Teff$, and metallicity [Fe/H] in the same way as for the \Kepler\ targets. To sum up, roughly 70\% of the spectroscopic estimates are from the APOGEE DR17 catalogue, 2\% are from GALAH DR3, and 28\% are from RAVE DR6.
The typical uncertainties on so obtained global parameters are $\sigma_{M} = 12\%$, $\sigma_{\mathrm{[Fe/H]}} = 0.06\,$dex, $\sigma_{\Teff} = 64$K, $\sigma_{\numax} = 9\%$. Our final sample is composed of 4099 stars,
including RGB and AGB stars, as well as stars leaving the clump phase. Some of them are shown in given bins of mass and metallicity in Fig.~\ref{fig:seism_HR_diag}.

\section{Stellar models}
\label{sec:models}


Evolutionary tracks and stellar models are derived with the release 12778 of the stellar evolution code Modules for Experiments in Stellar Astrophysics \citep[\mesa, ][]{2011ApJS..192....3P, 2013ApJS..208....4P, 2015ApJS..220...15P, 2018ApJS..234...34P, 2019ApJS..243...10P}. We computed a grid of stellar models with initial mass $M = [0.8,\ 0.9,\ 1.0,\ 1.1,\ 1.2,\ 1.5,\ 1.75,\ 2.0,\ 2.5]M_{\odot}$ and initial metallicity $\mathrm{[Fe/H]} = [-1.0, -0.5, -0.25, 0.0, 0.25]\,$dex. The initial fractional abundance of metals in mass was set following the solar chemical composition described in \cite{2009ARA&A..47..481A}. The treatment of convection is based on the mixing-length formalism presented in \cite{1965ApJ...142..841H}, which takes the opacity of the convective eddies into account. The initial helium abundance $Y_{0}$, the metallicity [Fe/H] and the mixing-length parameter $\alphaMLT$ were calibrated to reproduce the present solar luminosity, radius, and surface metal abundance. To this end, we adapted the \mesa\ test suite case \textit{simplex\_solar\_calibration} and took the $\log L$, $\log R$ and $Z/X$ terms\footnote{$Z/X = (Z/X)_{\odot}\ 10^{\mathrm{[Fe/H]}}$, where the solar value $(Z/X)_{\odot} = 0.0181$ is taken from \cite{2009ARA&A..47..481A}} into account in the $\chi^{2}$ value. We performed the solar calibration without microscopic diffusion. This gave us the solar-calibrated values $Y_{0} = 0.253$, $\mathrm{[Fe/H]} = 0$ (equivalently $Z_{0} = 0.0133$), and $\alphaMLT = 1.92$ at a solar age of 4.61 Gyrs 
\footnote{This value corresponds to the default solar age in \mesa, taken as the sum of the time spent on the MS starting on the ZAMS ($4.57 \times 10^{9}\,$yrs) and that spent on the PMS ($0.04 \times 10^{9}\,$yrs). 
It is larger than the value commonly adopted $\tau_{\odot} = 4.57 \times 10^{9}\,$yrs \citep[see, e.g., ][]{2007leas.book...45C}.
However, we do not account for the PMS in the calibration and it has been shown that adopting those two target solar ages does not impact the solar calibration significantly \citep[see Table~2 of][]{1990ApJ...360..727S}.}.
We do not assume any coupling of $Y$ and $Z$ through the $\Delta Y / \Delta Z$ helium-to-metal Galactic enrichment ratio, but rather explore different values of the couple $(Y, Z)$.  \\

We started from the \textit{1M\_pre\_ms\_to\_wd} test suite case and customised the physical ingredients to model stellar evolution up to the AGB. To follow chemical changes and the production of nuclear energy, we used a network of 32 nuclear reactions involving 23 stable or unstable species from $^1$H to $^{24}$Mg. The thermonuclear reaction rates are taken from NACRE \citep{1999NuPhA.656....3A} and CF88 \citep{1988ADNDT..40..283C}, with priority on NACRE rates when available. We took into account some updates to crucial reaction rates at evolved stages, such as $^{14}$N$(p,\gamma)^{15}$O \citep{2004A&A...420..625I} and triple-$\alpha$ \citep{2005Natur.433..136F}. 

Opacities are needed to compute the energy transport in regularly stratified regions, that is in radiative zones. At low temperatures ($\log T < 3.95$), we used the opacity tables from AESOPUS \citep{2009A&A...508.1539M} 
while at high temperatures ($\log T > 4.05$), we took  either the OPAL1 or OPAL2 opacity tables \citep{1996ApJ...464..943I}. The AESOPUS and OPAL1 tables are for the \citet{2009ARA&A..47..481A} solar mixture, while the OPAL2 tables allow to account for the metal mixture changes due to the C and O enhancements that result from He burning.
In the region around $\log T = 4.00 \pm 0.05$, we performed a blend as described by \citet{2011ApJS..192....3P} (see their Eq. (1)) between AESOPUS and OPAL1 tables. Furthermore, in the regions with C and O enhancements where the initial metallicity $Z_0$ is increased by an amount $\diff Z$ we used the OPAL2 opacity tables. A blend between OPAL1 and OPAL2 is made in the region where $\diff Z \in [0.001, 0.01]$.
Finally, the resulting opacity is combined with the electron conduction opacity, as prescribed in \cite{2007ApJ...661.1094C}.\\

We considered convective core overshooting during the main sequence following a step scheme \citep[e.g.][]{1975A&A....40..303M}. It means that the mixing region extends over a distance $d_{\mathrm{ov}}$ toward the surface from the boundary of convective instability in the core where the Schwarzschild condition $\gradrad > \gradad$ is fulfilled. If $H_{P} \leq R_{cc}$, where $H_{P}$ is the pressure scale height at the boundary of the convective core and $R_{cc}$ is the convective core radial thickness, then we keep $d_{\mathrm{ov}} = \alphaovH H_{P}$, otherwise $d_{\mathrm{ov}} = \alphaovH R_{cc}$. We adopted the overshooting parameter $\alphaovH = 0.2$ in the reference model. Similarly, we computed evolutionary tracks with and without convective core overshooting during core He-burning phase following two scenarii. Either the temperature gradient $\nabla_{T}$ is kept equal to the radiative gradient $\gradrad$ (usual overshooting scenario) or it is kept equal to the adiabatic temperature gradient $\gradad$ (penetrative convection scenario) in the overshooting region \citep{1991A&A...252..179Z}. The Schwarzschild convective border is in an unstable equilibrium in the sense that a small expansion of the convective core may make $\gradrad$ larger than $\gradad$ at the new border
\citep{1971Ap&SS..10..340C}. Adding core overshooting during the core He-burning phase induces a possible local minimum and maximum in $\gradrad$ in the extra-mixing region due to the increasing opacity that results from the transport of C and O elements in that region. Should this maximum increase above $\gradad$, a separate convective instability would occur in the overshooting region in absence of an appropriate treatment of semi convection. To address this issue, we allowed a partially mixed He-semiconvection region \citep{1971Ap&SS..10..355C} between the minimum of $\gradrad$ and the outer radiative zone following the diffusion scheme of \citet{1985A&A...145..179L} with the efficiency factor $\alpha\ind{sc} = 0.1$. In parallel, we followed the treatment proposed by \cite{2017MNRAS.469.4718B} to consider a stable convective border and suppress such spurious convective instabilities. This treatment consists in defining the convective border at the point where $\gradrad$ has reached its local minimum if the local maximum of $\gradrad$ created by adding He-core overshooting has increased over $\gradad$ in the extra-mixing region, otherwise the convective boundary is set at the point where $\gradrad = \gradad$. \\

We also explored the effect of envelope undershooting, which induces an extra-mixing region of extent $\alphaundersh H_{P}$ below the convective envelope into the radiative core. Similarly as convective core overshooting, we adopted a step scheme and applied envelope undershooting from the main sequence up to the AGB, with $\nabla_{T} = \gradrad$ in the extra-mixing region. The typical value for low-mass red giants $(M \leq 1.6 M_{\odot})$ at solar metallicity recommended by \cite{2018ApJ...859..156K}, i.e. $\alphaundersh = 0.3$ is considered. \\
Other transport processes arise in stellar interiors, such as thermohaline mixing and rotation-induced mixing. Thermohaline convection starts along the RGB in regions that are stable against convection (according to the Ledoux criterion) and where the molecular weight gradient becomes negative (i.e. $\nabla_{\mu} = \mathrm{d}\ln \mu / \mathrm{d}\ln P < 0$) between the H-burning shell surrounding the degenerate core and the convective envelope. This composition gradient inversion is induced by the $^{3}\mathrm{He}(^{3}\mathrm{He},2p)^{4}\mathrm{He}$ reactions that take place around the H-burning shell \citep{1972ApJ...172..165U, 2006Sci...314.1580E, 2008ApJ...677..581E, 2007A&A...467L..15C}. Here, thermohaline mixing is treated in a diffusion approximation based on the work of \cite{1980A&A....91..175K}, where the corresponding diffusion coefficient reads \citep{2013ApJS..208....4P} \\

\begin{equation}
    \label{eq:diffusion_thermohaline}
    \coefdiffthline = \alphath \frac{3K}{2\rho c_{P}}\frac{B}{\nabla_{T} - \gradad}.
\end{equation}
In the previous equation, $K$ is the radiative conductivity, $c_{P}$ is the specific heat at constant pressure, 
$\alphath$ is the efficiency parameter for the thermohaline mixing, and 

\begin{equation}
    \label{eq:B_expression}
    B = -\frac{1}{\chi_{T}} \sum_{i = 1}^{N-1}{\left( \frac{\partial \ln P}{\partial \ln X_{i}} \right)_{\rho,T,\{X_{j\neq i}\}}\frac{\diff \ln X_{i}}{\diff \ln P}},
\end{equation}
where $\chi_{T} = (\partial \ln P / \partial \ln T)_{\rho}$, and $X_{i}$ represents the mass fraction of atoms of species $i$ in the $N$-component plasma. The species $j$ is eliminated in the sum so that the constraint $\sum_{i = 1}^{N-1}{X_{i}} + X_{N} = 1$ is fulfilled.
We adopted $\alphath = 2$, which corresponds to the prescription of \cite{1980A&A....91..175K} where blobs of size $L$ diffuse while travelling over a mean free path $L$ before dissolving. We checked that the extent of the extra mixing regions caused by thermohaline convection is in agreement with the one obtained by \cite{2010A&A...521A...9C}. Especially, we verified that the extra mixing region is large enough to connect the H-burning shell and the convective envelope during the core He-burning phase for stars with a mass below $1.5 M_{\odot}$. \\

\begin{table*}[t]
\caption{Number of stars per mass and metallicity bins}
\begin{center}
\begin{tabular}{rrrrrr}
\hline
\hline
 & [Fe/H] (dex) $\in [-1.0, -0.5]$ & $[-0.5, -0.25]$ & $[-0.25, 0.0]$ & $[0.0, 0.25]$ \\
 
 $M/M_{\odot} \in $ & & & & \\
\hline
 $[0.6, 0.9]$ & 122$\qquad$ & 174$\qquad$ & 204$\ \ \ \ $ & 107$^{(*)}$ $\ $\\
\hline
 $[0.9, 1.2]$ & 130$\qquad$ & 344$\qquad$ & 432$\ \ \ \ $ & 213$\ \ \ \ \ $ \\
\hline
 $[1.2, 1.5]$ & 108$\qquad$  & 322$\qquad$ & 426$\ \ \ \ $ & 244$\ \ \ \ \ $ \\
\hline
 $[1.5, 2.5]$ & 143$\qquad$ & 346$\qquad$ & 518$\ \ \ \ $ & 266$\ \ \ \ \ $ \\
  \hline
  \label{Table:stars_per_bin}
\end{tabular}
\\
\end{center}
\textbf{Notes:} $^{(*)}$ refers to mass and metallicity bins for which the fit could not converge
\end{table*}

We investigated the effects of rotation on the AGBb location since rotation is known to impact lifetimes, surface abundances and evolutionary fates. Rotation is treated in 1D in the shellular approximation \citep[e.g.][]{1997A&A...321..465M}. Further information can be found in \cite{2013ApJS..208....4P} to learn more about how rotation is implemented in \mesa\ where it is treated as a diffusive process. We took rotation into account from the zero-age main sequence (ZAMS), as often done in stellar evolution codes \citep[e.g.][]{1989ApJ...338..424P}, up to the terminal-age main sequence (TAMS). First, we implemented rotation up to the early AGB phase, but it made the evolutionary track noisy at the AGBb without modifying its position. Therein, we only kept rotation during the main sequence. The rotation rate gradually reaches the maximum value $\Omega_{\mathrm{ZAMS}}/\Omegacrit = 0.3$, where $\Omegacrit$ is the surface critical angular velocity for the star to be dislocated, which is the typical rotation rate motivated by observations of B stars \citep{2010ApJ...722..605H}. For a 2$M_{\odot}$ star, the evolutionary track that includes the rotation rate $\Omega_{\mathrm{ZAMS}}/\Omegacrit = 0.3$ during the main sequence is equivalent to that including a H-core overshooting $\alphaovH \approx 0.25$. We checked that the evolution of the surface rotation rate so obtained along the main sequence is similar to that obtained in \cite{2012A&A...537A.146E}. We only studied rotating models with $M \geq 1.5 M_{\odot}$ since magnetic braking is not included in \mesa, which does not allow to reproduce slow rotation rates of low-mass stars \citep{1988ApJ...333..236K}. \\
Rotation induces both chemical and angular momentum transports through instabilities that are treated in a diffusion approximation \citep{1978ApJ...220..279E, 1989ApJ...338..424P, 2000ApJ...528..368H}. In our models, we included six equally weighted instabilities induced by rotation which are dynamical shear, Solberg-H{\o}iland, secular shear, Goldreich-Schubert-Fricke instabilities, Eddington-Sweet circulation, and Tayler-Spruit dynamo. Then, each diffusion coefficient associated to those rotationally induced instabilities is added to the diffusion coefficient in absence of rotation. On top of this resulting diffusion coefficient, two free parameters need to be fixed in diffusion equations: the factor $f_{c}$ that scales the efficiency of composition mixing relatively to that of the angular momentum transport, and the factor $f_{\mu}$ that encodes the sensitivity of rotational mixing to the mean molecular weight gradient. Typical values from \cite{2000ApJ...528..368H} such as $f_{c} = 1/30$ and $f_{\mu} = 0.05$ are adopted. \\

We took a grey atmosphere with an Eddington $T(\tau)$ relation. We defined the outermost meshpoint of the models as the layer where the optical depth $\tau$ verifies $\tau = 2/3$, which is at the limit of the photosphere. Another important parameter that impacts the fate of stars is the mass-loss rate. We used Reimers' prescription \citep{1975MSRSL...8..369R} 

\begin{equation}
\label{eq:Reimers_mass_loss}
\dot{M}_{R} = -4\times 10^{-13}\ \eta_{R} \frac{L}{L_{\odot}} \frac{R}{R_{\odot}}\left(\frac{M}{M_{\odot}}\right)^{-1} M_{\odot}.\mathrm{yr}^{-1}
\end{equation}
from the RGB up to the core He-burning phase, where $\eta_{R}$ is the Reimers' scaling factor that we take equal to $\eta_{R} = 0.3$ \citep{2021A&A...645A..85M}. On the AGB, we use the Bl\"ocker's prescription \citep{1995A&A...297..727B}
\small
\begin{equation}
\label{eq:Blocker_mass_loss}
\dot{M}_{B} = - 1.93\times 10^{-21} \eta_{B}\ \left(\frac{M}{M_{\odot}}\right)^{-3.1}\frac{R}{R_{\odot}} \left(\frac{L}{L_{\odot}}\right)^{3.7} M_{\odot}.\mathrm{yr}^{-1},
\end{equation}
\normalsize
where $\eta_{B}$ is the Bl\"ocker's scaling factor taken equal to $\eta_{B} = 0.1$. In both Reimers' and Bl\"ocker's prescriptions, $L$, $R$ and $M$ are expressed in solar units. \\
The screening factors were computed with the implementation of \cite{2007PhRvD..76b5028C} for weak and strong screening conditions. We kept the default coverage of the equation of state in the $\log\rho-\log T$ plane, they are summarised in Fig.~50 of \cite{2019ApJS..243...10P}. \\


\section{Characterisation of the AGBb}
\label{sec:method}

\subsection{Observations}

Herebefore we found that the AGBb manifests as a local excess of stars on top of a background composed of RGB and AGB stars. In order to infer the AGBb location in $\numax$ and $\Teff$, in the way \citet{2018ApJ...859..156K} proceeded to characterise the RGBb, we adopted the statistical mixture model presented in \cite{2010arXiv1008.4686H}. This approach is a statistical framework where the data set is assumed to be multimodal, i.e. with several regions of high probability separated by regions of low probability. In this situation, we modelled the data with a mixture of several components, where each data point belongs to one of these components. This allowed us to use multiple models to fit our data set. We distinguished the inliers, which are stars belonging to the AGBb overdensity and the outliers, which are stars that belong to the RGB/AGB background and do not lie in the AGBb phase. The fit was performed using the \python\ module \emcee, which is an affine invariant Markov Chain Monte Carlo (MCMC) ensemble sampler \citep{2013PASP..125..306F}. The likelihood function is defined as

\begin{figure*}[htbp]
	\includegraphics[width=1.0\linewidth]{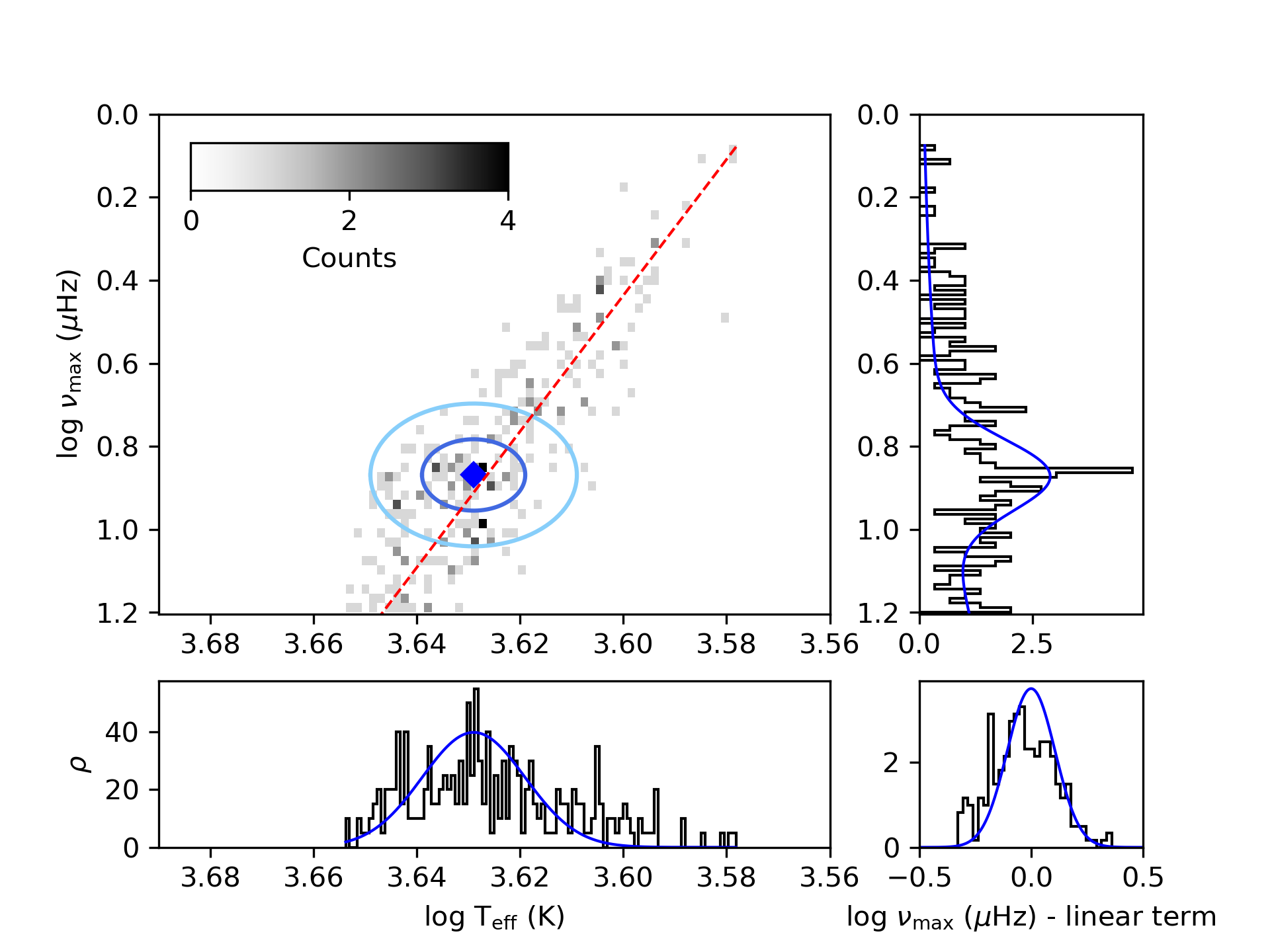}
	\caption{Probability distribution functions of our data set in the $\log\Teff - \log\numax$  plane, in the bins $M\in [0.9, 1.2]M_{\odot}$ and [Fe/H] $\in [-0.25, 0.0]\,$dex. Upper left panel: 2D histogram where the AGBb is located by a blue diamond. Dark blue and light blue ellipses correspond to the 1$\sigma$ and 2$\sigma$ regions of the bivariate Gaussian, respectively. The red dashed line reproduces the linear term belonging to the RGB/AGB background. Upper right panel: the normalised 1D histogram in $\log\numax$ is shown in black. The ordinate axis is the same as in the upper left panel. The blue line corresponds to the probability distribution function made of the Gaussian associated to the overdensity in $\log\numax$, multiplied by the rising exponential in $\log\numax$. Lower left panel: same label as in the upper right panel but in terms of $\log\Teff$. The abscissa axis is the same as in the upper left panel. The blue line shows the Gaussian associated to the overdensity in $\log\Teff$. Lower right panel: difference between $\log\numax$ and $\abg \log\Teff + \bbg$. The blue line illustrates the normal distribution around the linear term.
	}
	\label{fig:fit_AGBb}
\end{figure*}

\small
\begin{equation}
\label{eq:likelihood_mixture_model}
\mathcal{L} = (1 - \Pbg) \fbiv(\log \Teff, \log \numax) + \Pbg \fbg(\log \Teff, \log \numax),
\end{equation}
\normalsize
where $\fbiv$ describes the AGBb foreground with a bivariate normal distribution function, $\fbg$ describes the RGB/AGB background with the product of a normalised rising exponential in $\log \numax$ and a linear term with a normally distributed scatter, and $\Pbg$ is the mixture model weighting factor that gives the probability for a star to belong to the RGB/AGB background. The foreground and background probability distribution functions are

\begin{equation}
\label{eq:foreground_prob_func}
\fbiv(x_{1}, x_{2}) = \frac{1}{2\pi \sigma_{1}\sigma_{2}\sqrt{1 - \rho_{12}^{2}}}\mathrm{e}^{-\frac{z}{2(1 - \rho_{12}^{2})}} ,
\end{equation}

$$
\mathrm{with}\ z = \frac{(x_{1} - \mu_{1})^{2}}{\sigma_{1}^{2}} + \frac{(x_{2} - \mu_{2})^{2}}{\sigma_{2}^{2}} - \frac{2\rho_{12}(x_{1} - \mu_{1})(x_{2} - \mu_{2})}{\sigma_{1}\sigma_{2}}
$$
and

\begin{equation}
\label{eq:background_prob_func}
\fbg(x_{1}, x_{2}) = \frac{1}{\sqrt{2\pi} \sigmabg}\mathrm{e}^{-\frac{(x_{2} - (\abg x_{1} + \bbg))^{2}}{2\sigmabg^{2}}} \times \Aexp\ \mathrm{e}^{\cbg \numax},
\end{equation}
respectively. In the previous equations, $x_{1} = \log\Teff$ and $x_{2} = \log\numax$, $\mu_{1}$ and $\mu_{2}$ are the AGBb locations in $\log \Teff$ and $\log \numax$, respectively, $\sigma_{1}$ and $\sigma_{2}$ are the AGBb standard deviations in $\log \Teff$ and $\log \numax$, respectively, $\rho_{12}$ is the correlation of the bivariate Gaussian, $\abg$ and $\bbg$ are the linear coefficients, $\sigmabg$ is the standard deviation of the normal distribution of the linear term, $\cbg$ and $\Aexp$ are the coefficient and the normalisation factor of the exponential term, respectively. \\
Given the small amplitude of the AGBb overdensity in $\log \numax$ (see Fig.~\ref{fig:fit_AGBb}), $\sigma_{2}$ failed to converge when we used the fitting method described above. Consequently, we estimated $\sigma_{2}$ separately by fitting the AGBb overdensity with a normal distribution function in the 1D histogram of $\log \numax$. We took this estimate and kept it fixed during the MCMC process. Then, the posterior probability distributions of our set of 9 free parameters, which are $\mu_{1}$, $\mu_{2}$, $\sigma_{1}$, $\rho_{12}$, $\abg$, $\bbg$, $\sigmabg$, $\cbg$, and $\Pbg$, were visualised with the \python\ module \corner\ \citep{2016JOSS....1...24F}. First, the guess values of the bivariate Gaussian and the exponential term were extracted from the 1D histograms in log $\numax$ and log $\Teff$ and those of the linear term were obtained from the 2D histogram, as seen in Fig.~\ref{fig:fit_AGBb}. Then, parameters were left free to vary according to a uniform prior probability distribution. \\

We performed this fitting method in the $\log \Teff - \log \numax$ plane, in restricted bins of mass and metallicity, which are $M\in [0.6, 0.9]$, $[0.9, 1.2]$, $[1.2, 1.5]$, $[1.5, 2.5]\, M_{\odot}$ and [Fe/H] $\in [-1.0, -0.5]$, $[-0.5, -0.25]$, $[-0.25, 0.0]$, $[0.0, 0.25]$  dex. The bins are wider at high mass and low metallicity to include enough stars and hence ensure the free parameters to converge. The number of stars per bin is shown in Table~\ref{Table:stars_per_bin}. We show in Fig.~\ref{fig:fit_AGBb} the results for the bin $M \in [0.9, 1.2] M_{\odot}$ and [Fe/H] $\in [-0.25, 0.0]$ dex.

\subsection{Models}
\label{sec:method_subsec_models}

To extract the probability for a star to lie in a given bin of $\numax$ and $\Teff$ along its evolutionary track, we computed the inverse of the evolution speeds $\diff \tau/\diff \numax$ and $\diff \tau/\diff \Teff$, where $\tau$ is the stellar age. Then, we integrated $\diff \tau/\diff \numax$ and $\diff \tau/\diff \Teff$ over $\numax$ and $\Teff$, respectively. This gives us the fractional time that is spent in a given bin of $\numax$ and $\Teff$, respectively. We used this fractional time as a proxy of the probability distribution for a star to lie in a given bin of $\numax$ and $\Teff$. The procedure was repeated for each pair of mass and metallicity in our grid of stellar models presented in Sect.~\ref{sec:models}. We summed the probability distributions of all pairs of mass and metallicity lying in the considered bin of mass and metallicity, and we normalised the resulting probability distribution. Because the grid of stellar models is discontinuous compared to observations and because the probability distributions are narrow at the turning-backs of the AGBb (i.e. where the quantities $\diff \tau/\diff \numax$ and $\diff \tau/\diff \Teff$ change sign), we convolved the resulting probability distribution by a normal one. Eventually, the maximum of the convolved probability distribution was interpreted as the AGBb location. \\

\begin{figure*}[htbp]
	\begin{minipage}{1.\linewidth}  
		\rotatebox{0}{\includegraphics[width=0.5\linewidth]{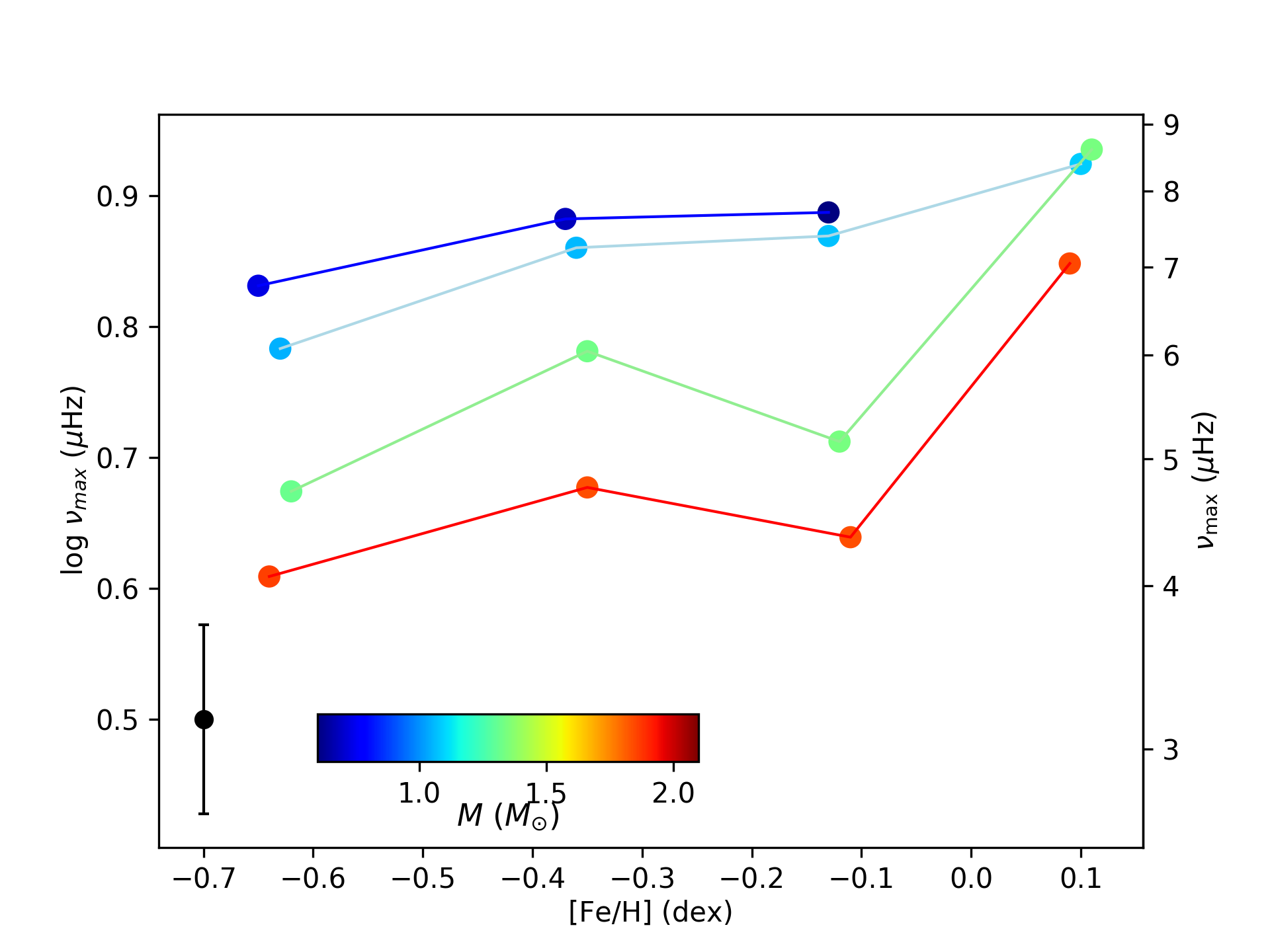}}
		\rotatebox{0}{\includegraphics[width=0.5\linewidth]{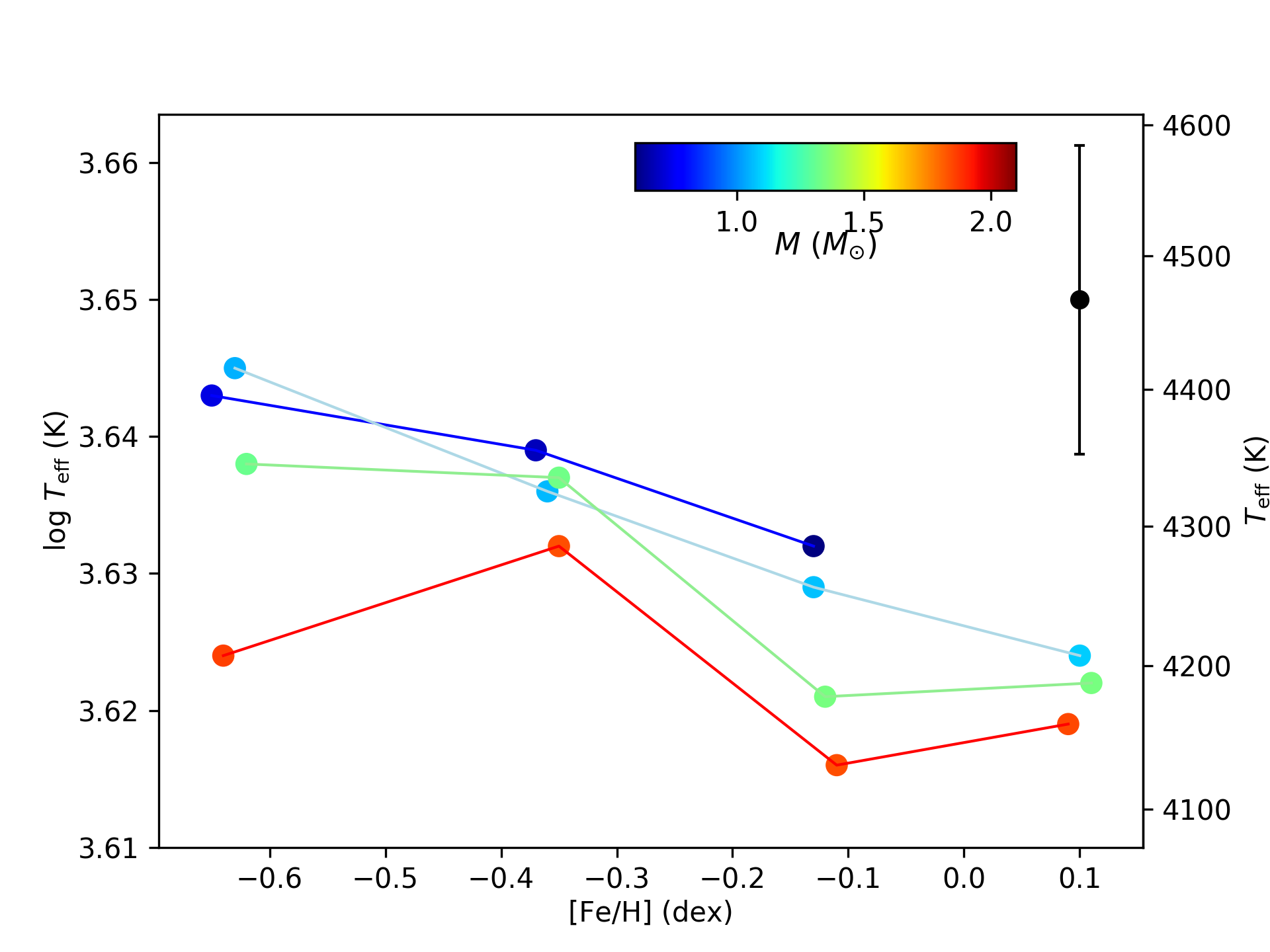}}
	\end{minipage}
	\caption{Location of the AGBb in $\log \numax$ (left) and in $\log \Teff$ (right) from observations, as a function of the metallicity [Fe/H]. The AGBb occurrence is marked by dots and the stellar mass is colour-coded. AGBb locations obtained in the same bin of mass $M\in [0.6, 0.9]$, $[0.9, 1.2]$, $[1.2, 1.5]$, $[1.5, 2.5]M_{\odot}$ are connected by dark blue, light blue, light green, and red lines, respectively. Mean error bars on the location of the AGBb in $\log \numax$ and in $\log \Teff$ are shown in black. Data in the bin ($M \in [0.6, 0.9]M_{\odot}$, [Fe/H]$\in [0, 0.25]\,$dex) are not shown because there are not enough stars to perform the statistical mixture model.
	}
	\label{fig:loc_AGBb_observations}
\end{figure*}

With the aim to investigate the potential of the AGBb to constrain physical processes in stellar interiors, we attempted to make the observed AGBb location match the expected AGBb one as well as possible by comparing the 1D histograms of $\log \numax$ and $\log \Teff$ from observations with the corresponding probability distributions derived from stellar models. To this end, we defined a reference model and varied the stellar parameters to explore their impact on the expected AGBb location in $\log \numax$ and $\log \Teff$. The results are presented in the following section.
	
\section{Results}
\label{sec:results}

\begin{figure}[htbp]
\centering
	\begin{minipage}{1.\linewidth}  
		\rotatebox{0}{\includegraphics[width=0.98\linewidth]{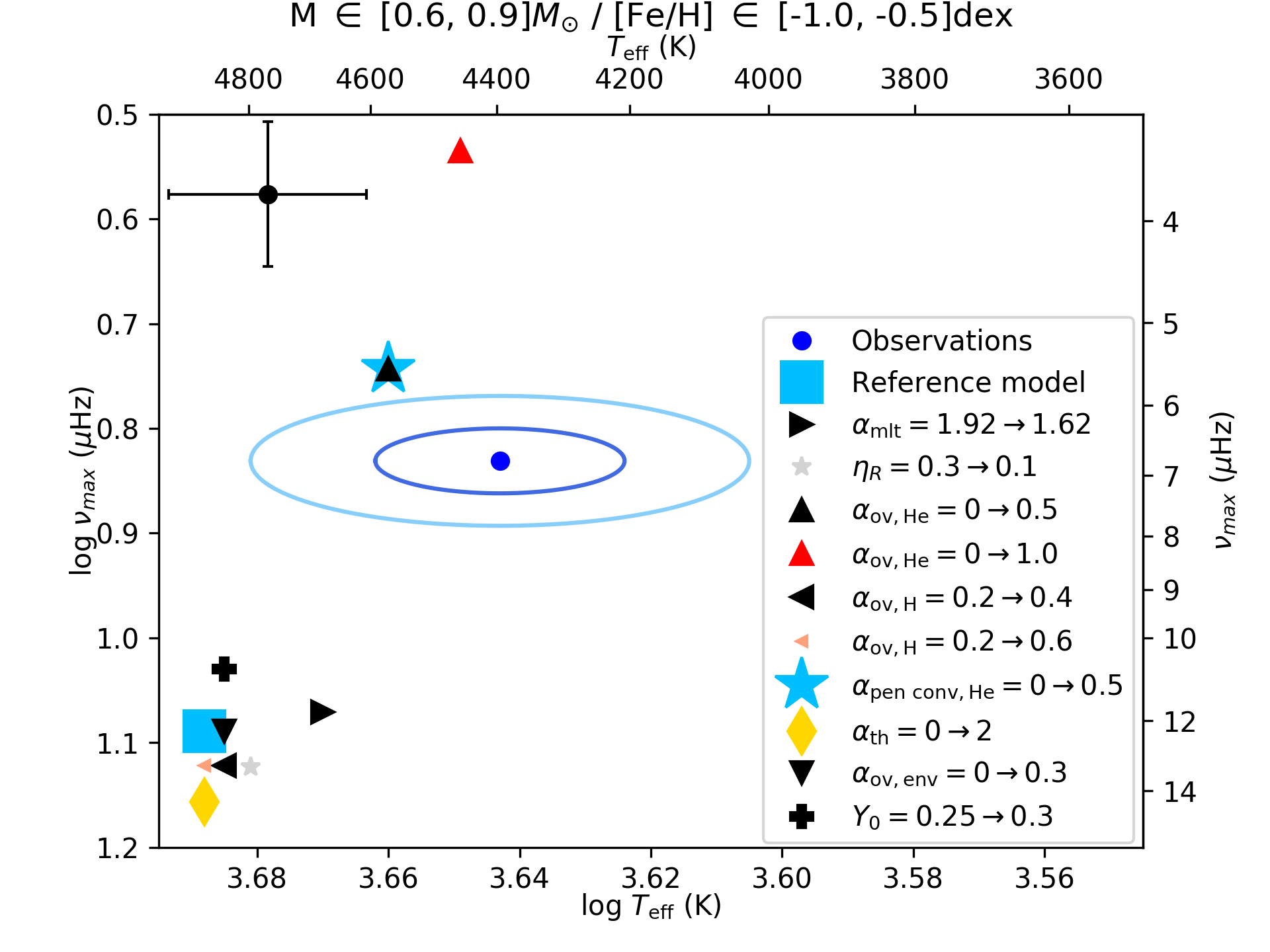}}
	\end{minipage}
	\begin{minipage}{1.\linewidth}  
		\rotatebox{0}{\includegraphics[width=0.98\linewidth]{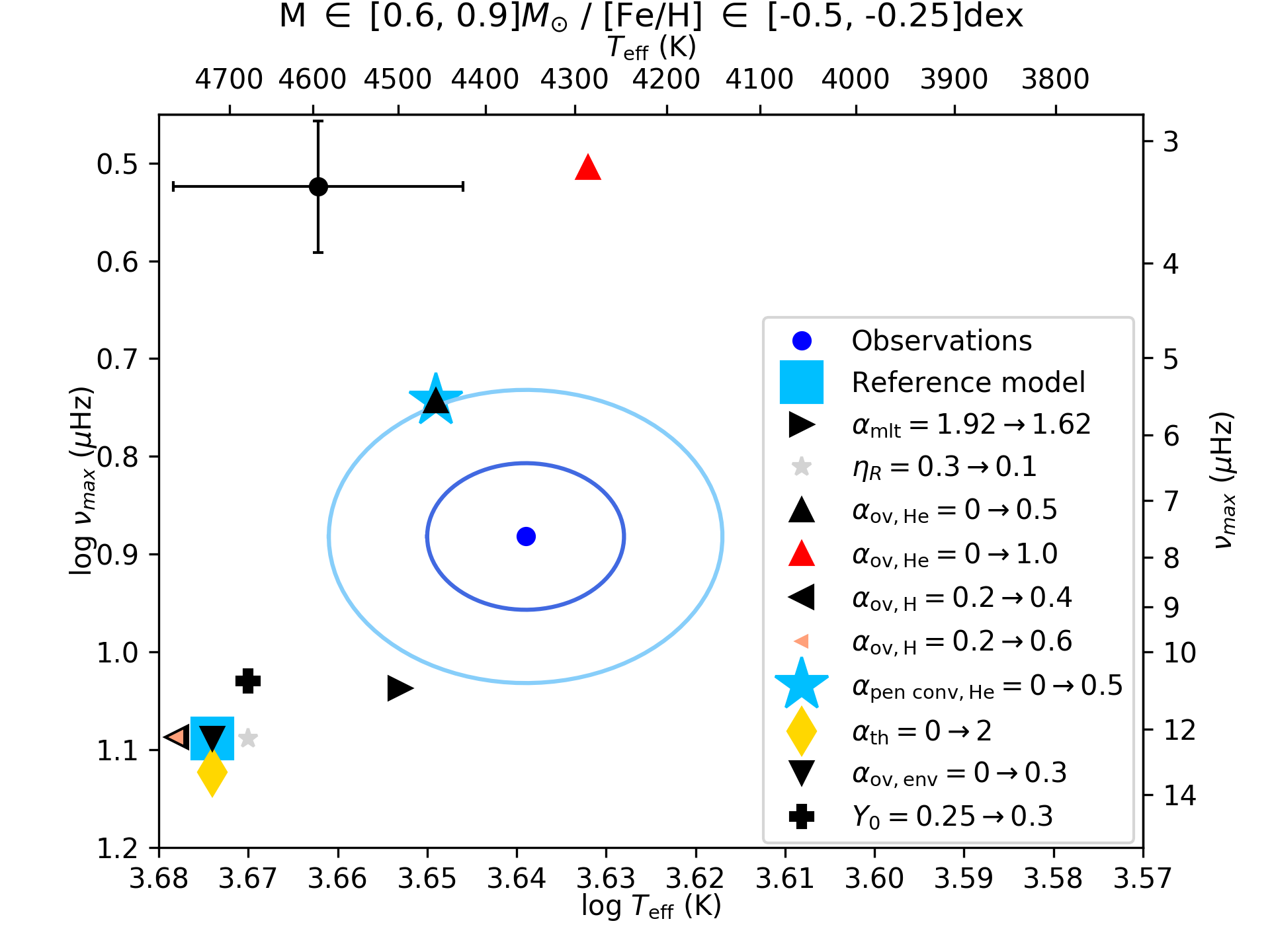}}
	\end{minipage}
	\begin{minipage}{1.\linewidth}  
		\rotatebox{0}{\includegraphics[width=0.98\linewidth]{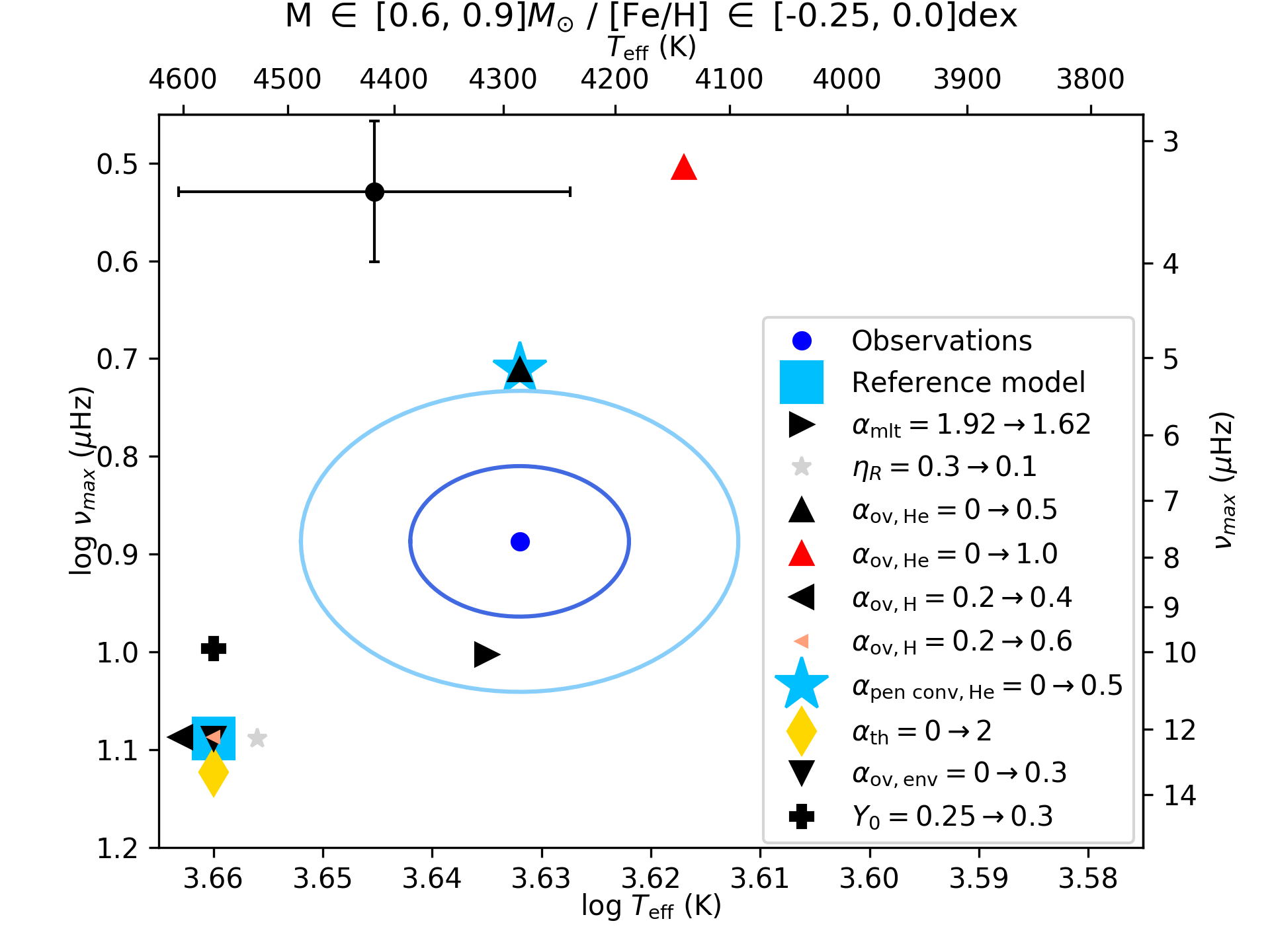}}
	\end{minipage} \\
	\caption{\small{Location of the AGBb in the plane $\log \Teff - \log \numax$ in the mass bin $M \in[0.6,0.9]M_{\odot}$ and metallicity bins [Fe/H] $\in [-1.0, -0.5]$, $[-0.5, -0.25]$, $[-0.25, 0.0]$ dex. The metallicity bin [Fe/H]$\in [0.0, 0.25]\,$dex is missing because we have not enough stars to perform a statistical study. Observations are marked by blue dots, the dark blue and light blue ellipses correspond to the 1$\sigma$ and 2$\sigma$ regions, respectively. The reference model presented is shown by a light blue square. Other models are shown with different symbols listed in the labels, they have been obtained by individually changing the parameters of the reference model. The changes are indicated in the label of each panel. The black dot with errorbars indicates the mean uncertainty we have for all models. The uncertainty on the AGBb location in $\log \numax$ and $\log\Teff$ for each model is taken as the standard deviation of the Gaussian function that reproduces the overdensity caused by the AGBb in the 1D histograms. The numerical values are listed in Table~\ref{table:AGBb_location_0_6_0_9M} in Appendix~\ref{appendix:AGBb_location}. Ranges of the axes vary in the different panels.}
	}
	\label{fig:loc_AGBb_0_6_to_0_9M}
\end{figure}

\begin{figure*}[htbp]
	\begin{minipage}{1.\linewidth}  
		\rotatebox{0}{\includegraphics[width=0.5\linewidth]{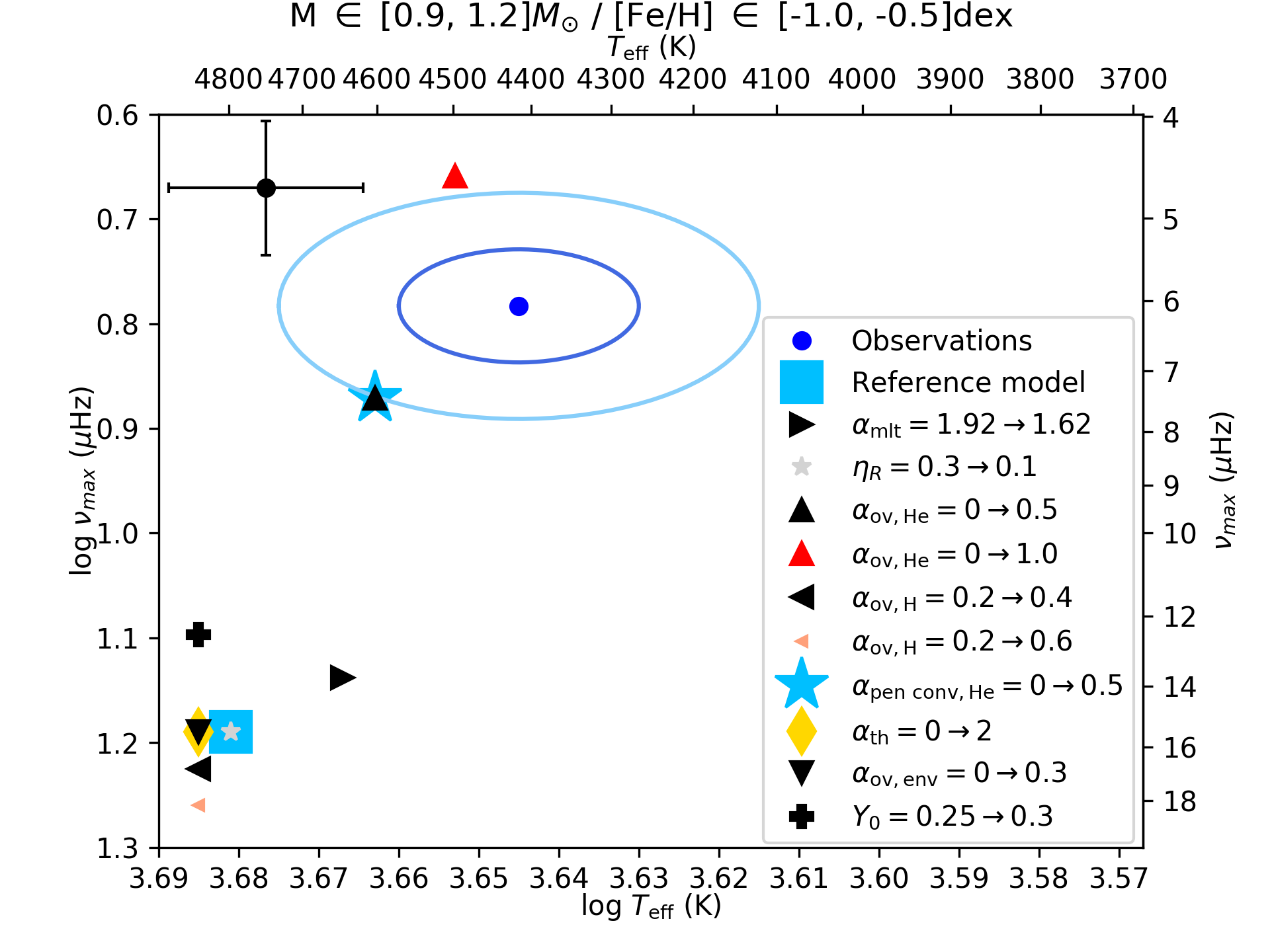}}
		\rotatebox{0}{\includegraphics[width=0.5\linewidth]{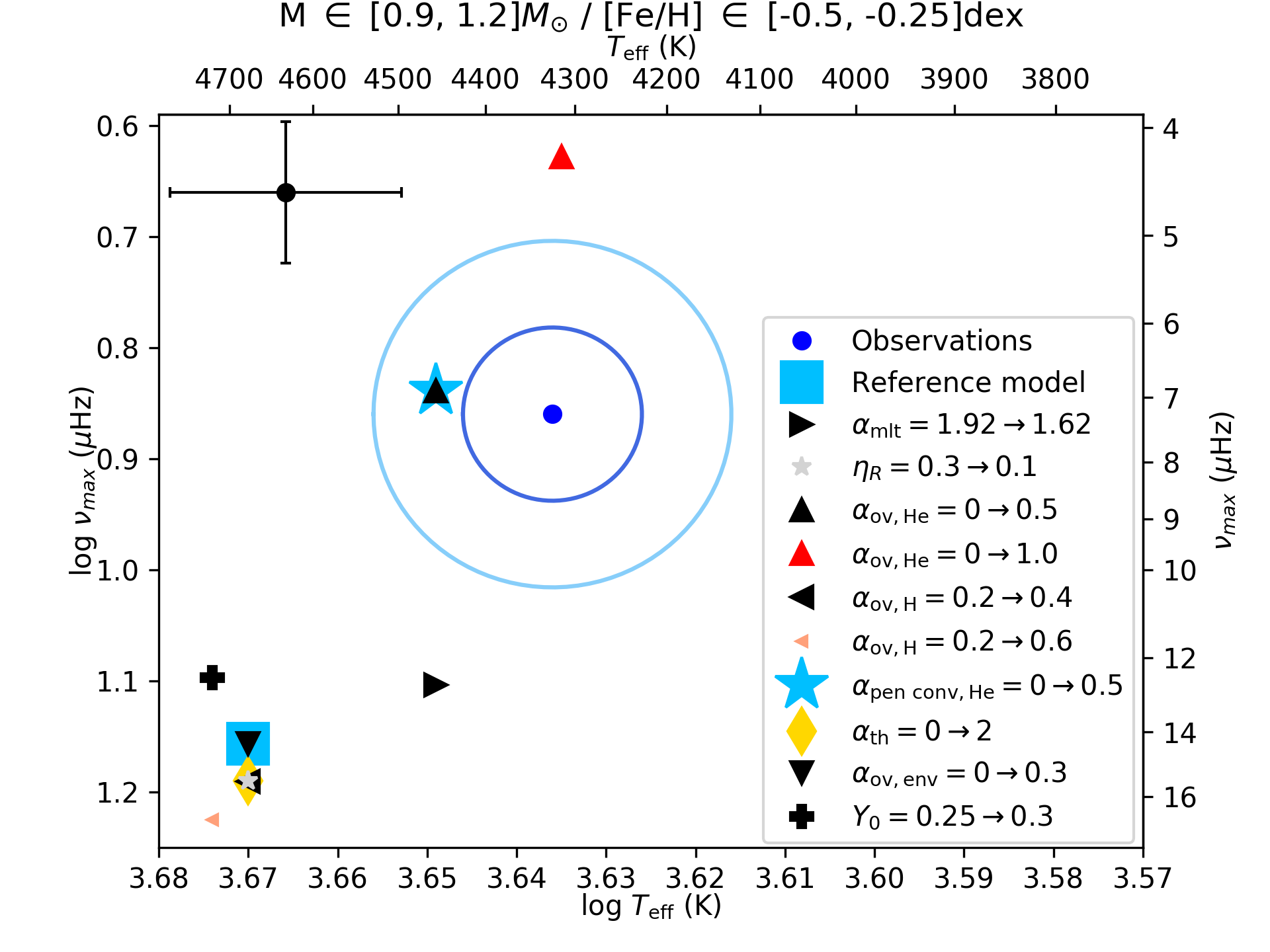}}
	\end{minipage}
	\begin{minipage}{1.\linewidth}  
		\rotatebox{0}{\includegraphics[width=0.5\linewidth]{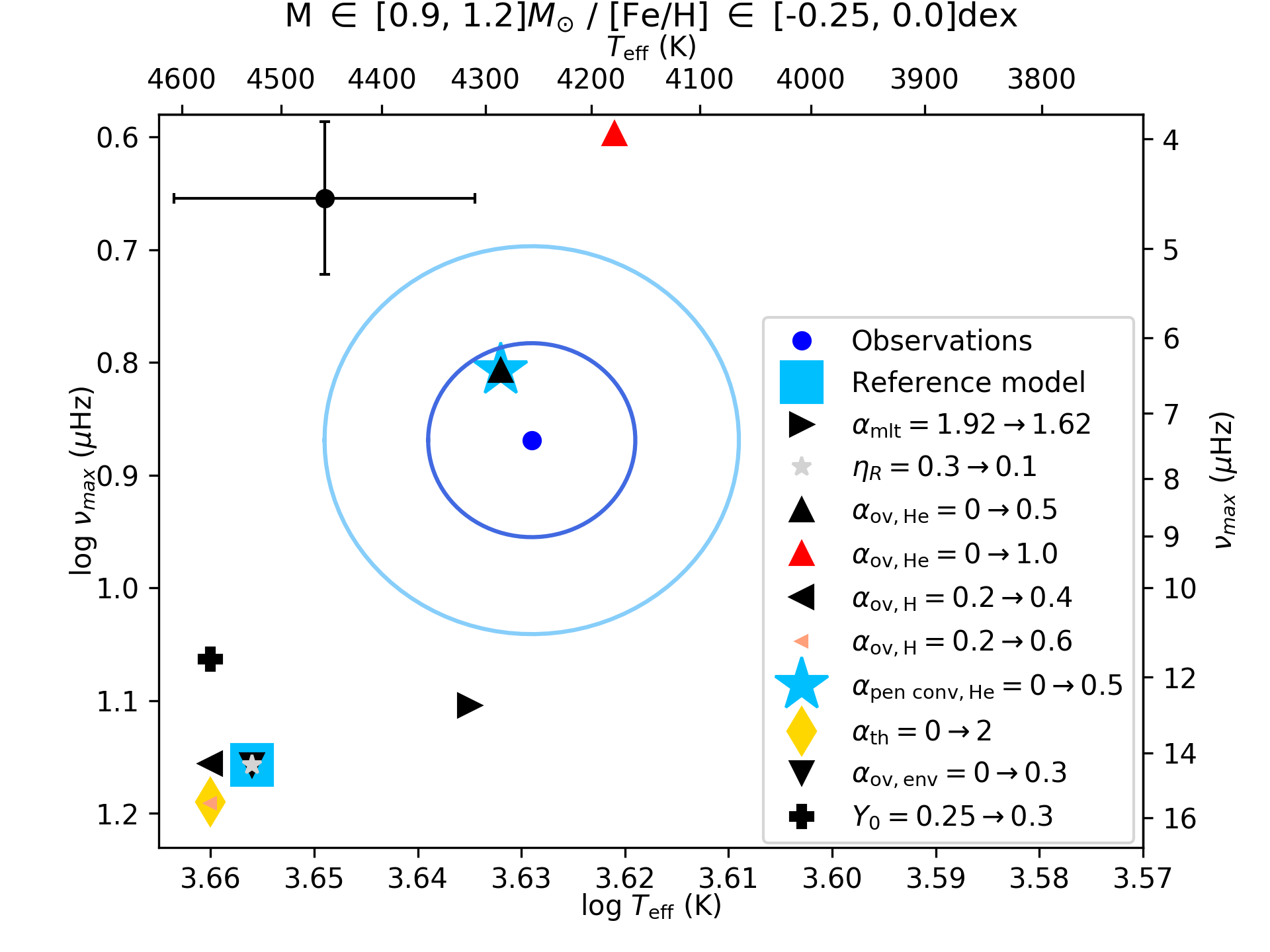}}
		\rotatebox{0}{\includegraphics[width=0.5\linewidth]{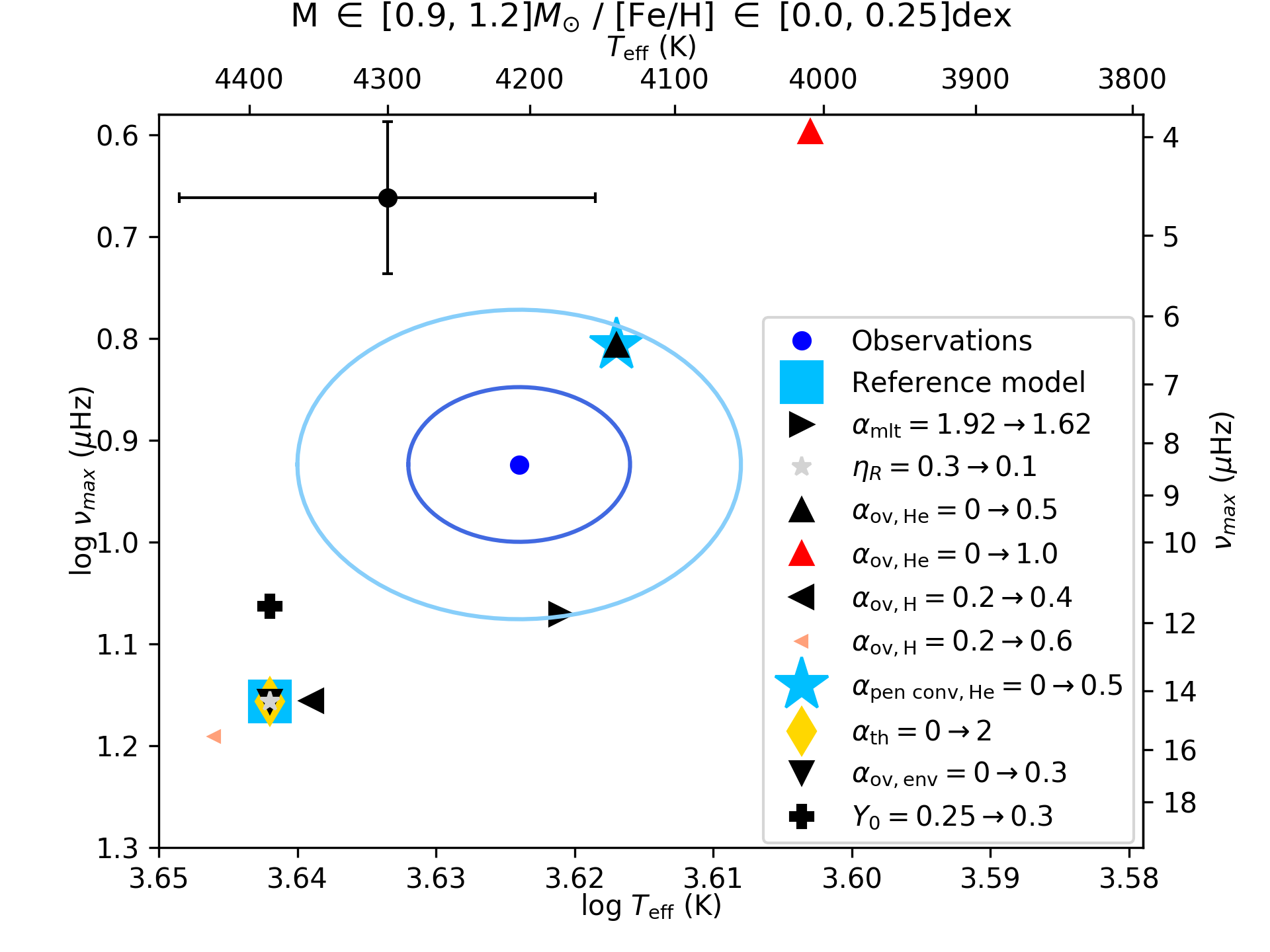}}
	\end{minipage}
	\caption{Same label as in Fig.~\ref{fig:loc_AGBb_0_6_to_0_9M}, but for the bins of mass $M \in[0.9,1.2]M_{\odot}$ and metallicity [Fe/H] $\in [-1.0, -0.5]$, $[-0.5, -0.25]$, $[-0.25, 0.0]$, $[0.0, 0.25]\,$dex. The numerical values are listed in Table~\ref{table:AGBb_location_0_9_1_2M} in Appendix~\ref{appendix:AGBb_location}. Ranges of the axes vary in the different panels.
	}
	\label{fig:loc_AGBb_0_9_to_1_2M}
\end{figure*}

\begin{figure*}[htbp]
\centering
	\begin{minipage}{1.\linewidth}  
		\rotatebox{0}{\includegraphics[width=0.5\linewidth]{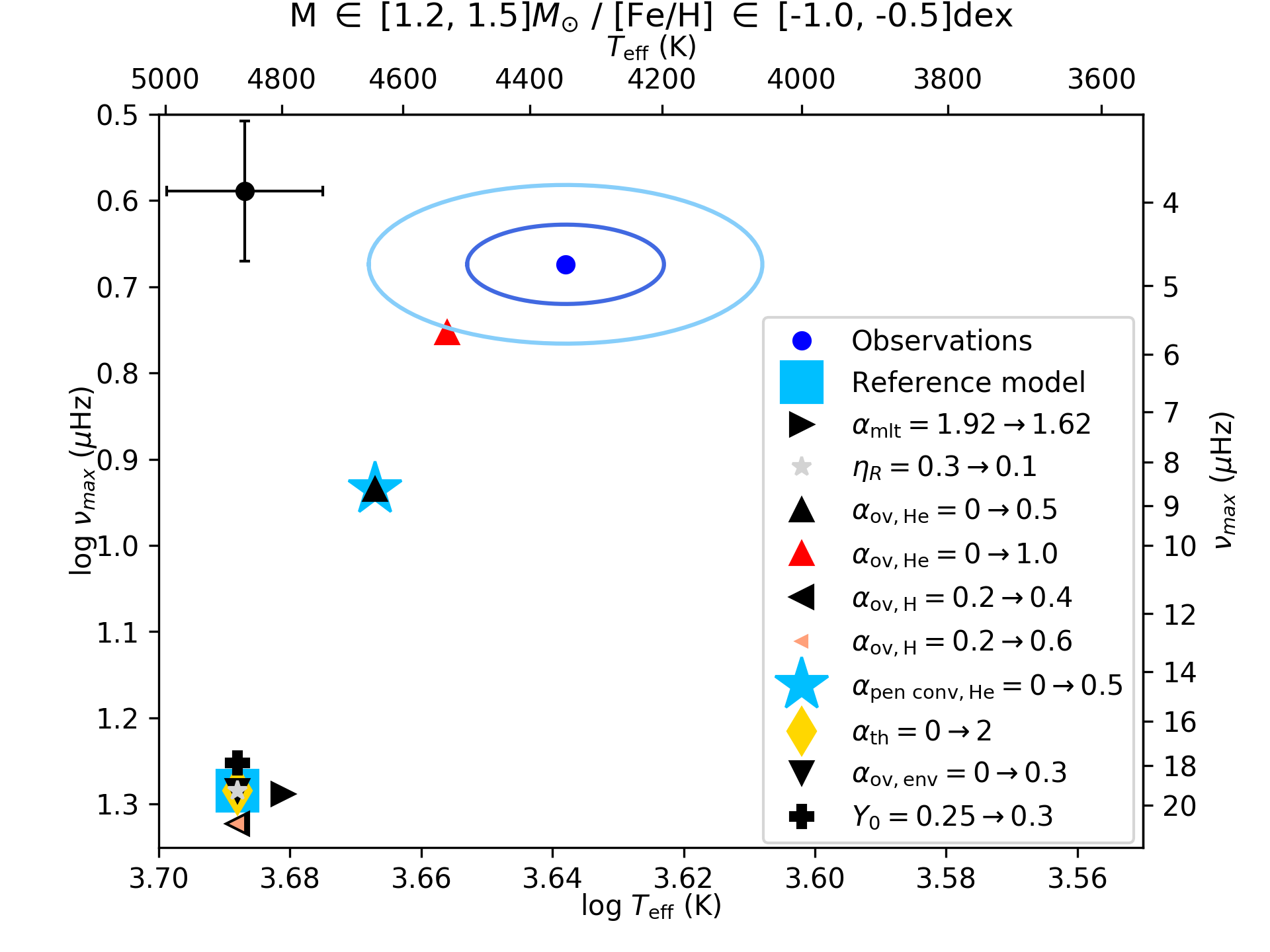}}
		\rotatebox{0}{\includegraphics[width=0.5\linewidth]{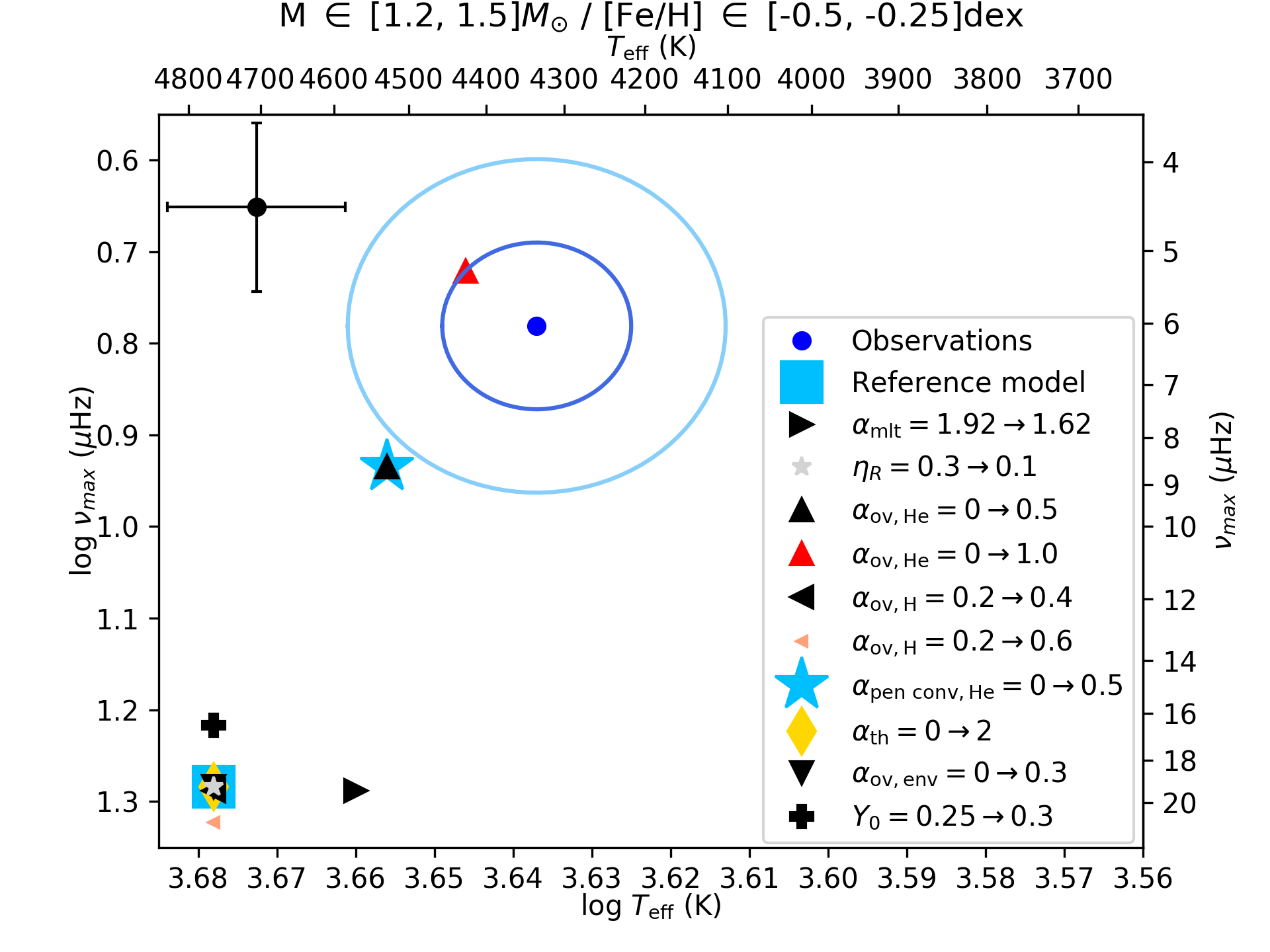}}
	\end{minipage}
	\begin{minipage}{1.\linewidth}  
		\rotatebox{0}{\includegraphics[width=0.5\linewidth]{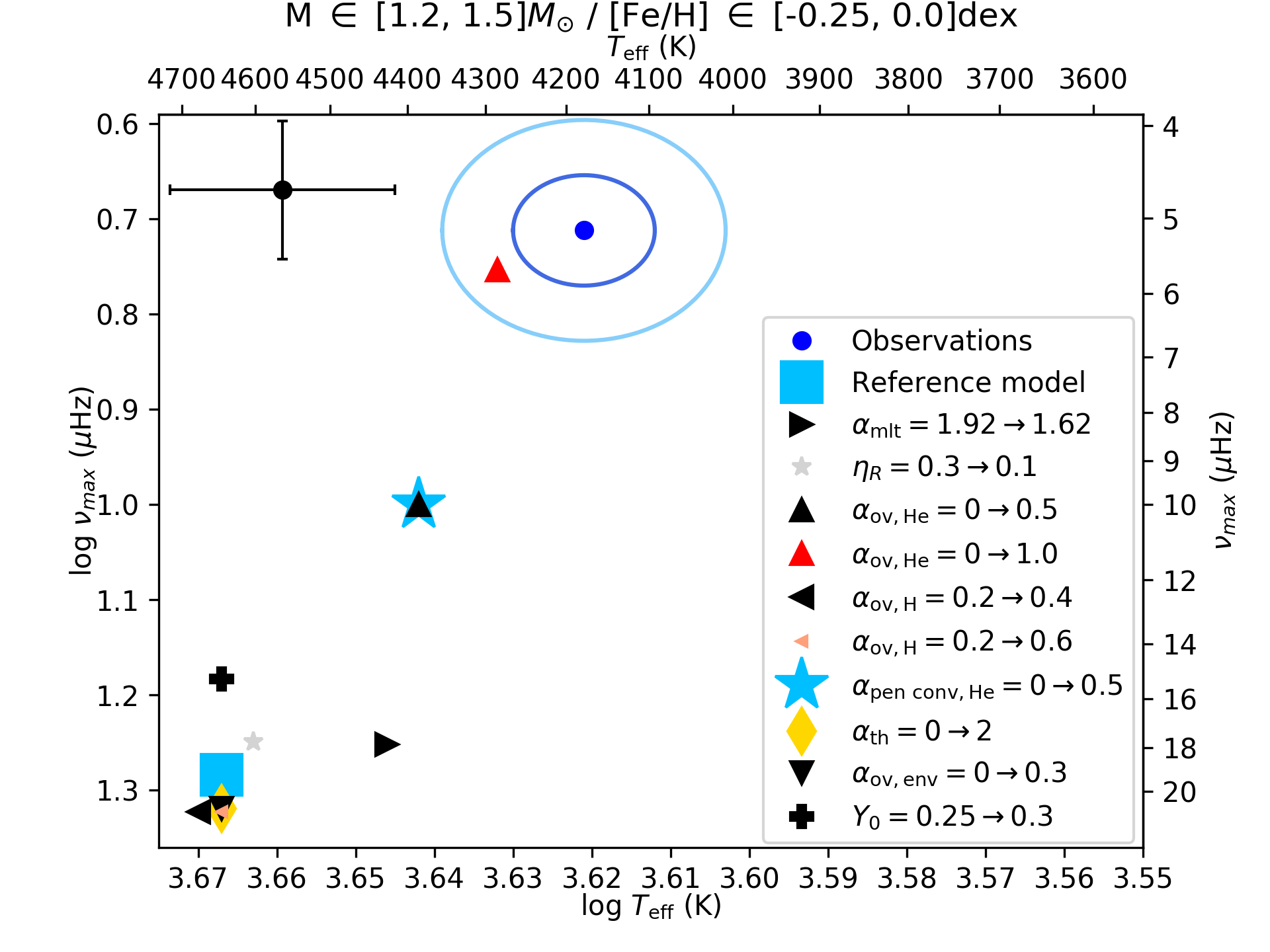}}
		\rotatebox{0}{\includegraphics[width=0.5\linewidth]{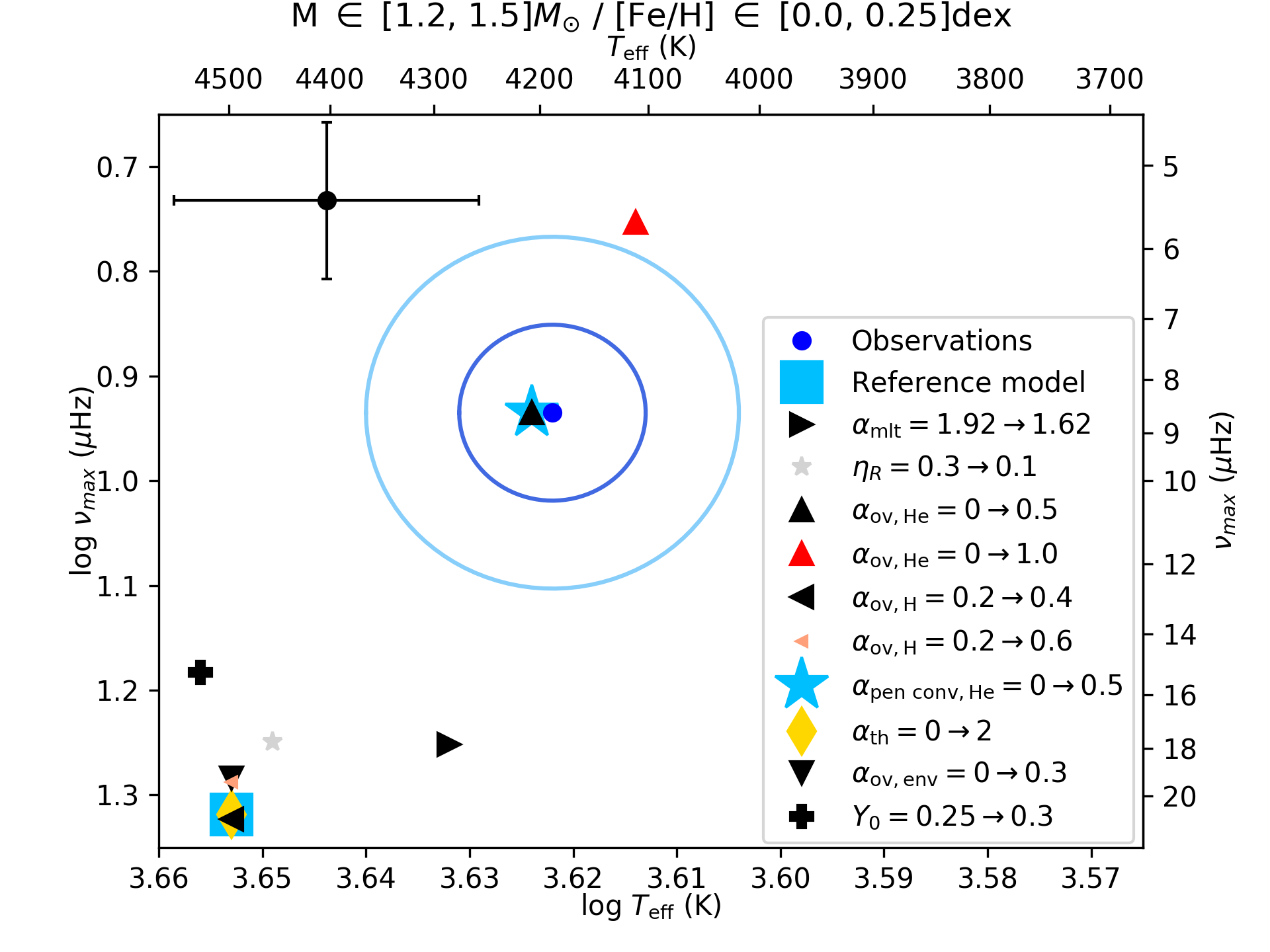}}
	\end{minipage}
	\caption{\small{Same label as in Fig.~\ref{fig:loc_AGBb_0_9_to_1_2M}, but for the bins of mass $M \in[1.2,1.5]M_{\odot}$ and metallicity [Fe/H] $\in [-1.0, -0.5]$, $[-0.5, -0.25]$, $[-0.25, 0.0]$, $[0.0, 0.25]\,$dex. The numerical values are listed in Table~\ref{table:AGBb_location_1_2_1_5M} in Appendix~\ref{appendix:AGBb_location}. Ranges of the axes vary in the different panels.}
	}
	\label{fig:loc_AGBb_1_2_to_1_5M}
\end{figure*}

\begin{figure*}[htbp]
\centering
	\begin{minipage}{1.\linewidth}  
		\rotatebox{0}{\includegraphics[width=0.5\linewidth]{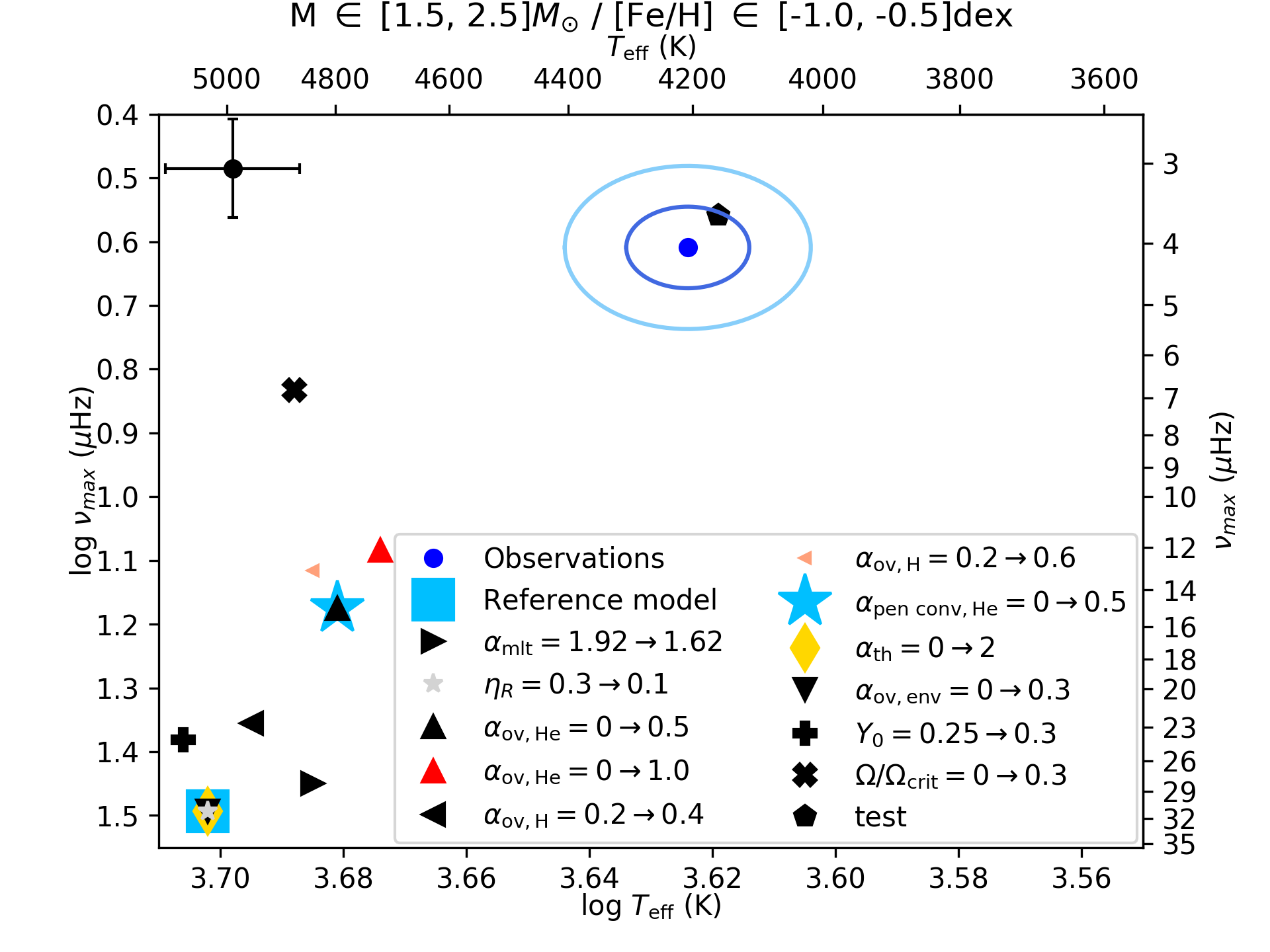}}
		\rotatebox{0}{\includegraphics[width=0.5\linewidth]{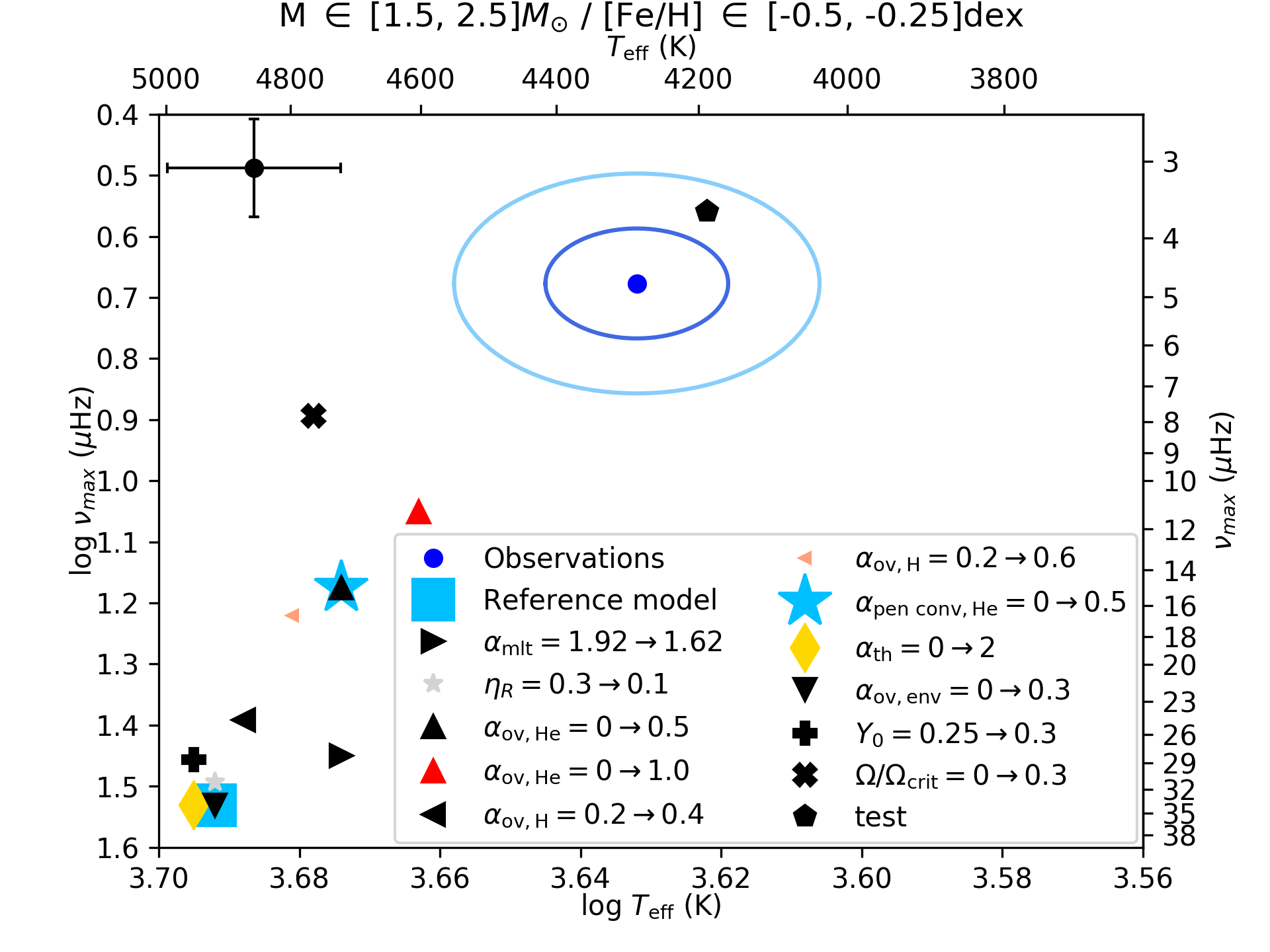}}
	\end{minipage}
	\begin{minipage}{1.\linewidth}  
		\rotatebox{0}{\includegraphics[width=0.5\linewidth]{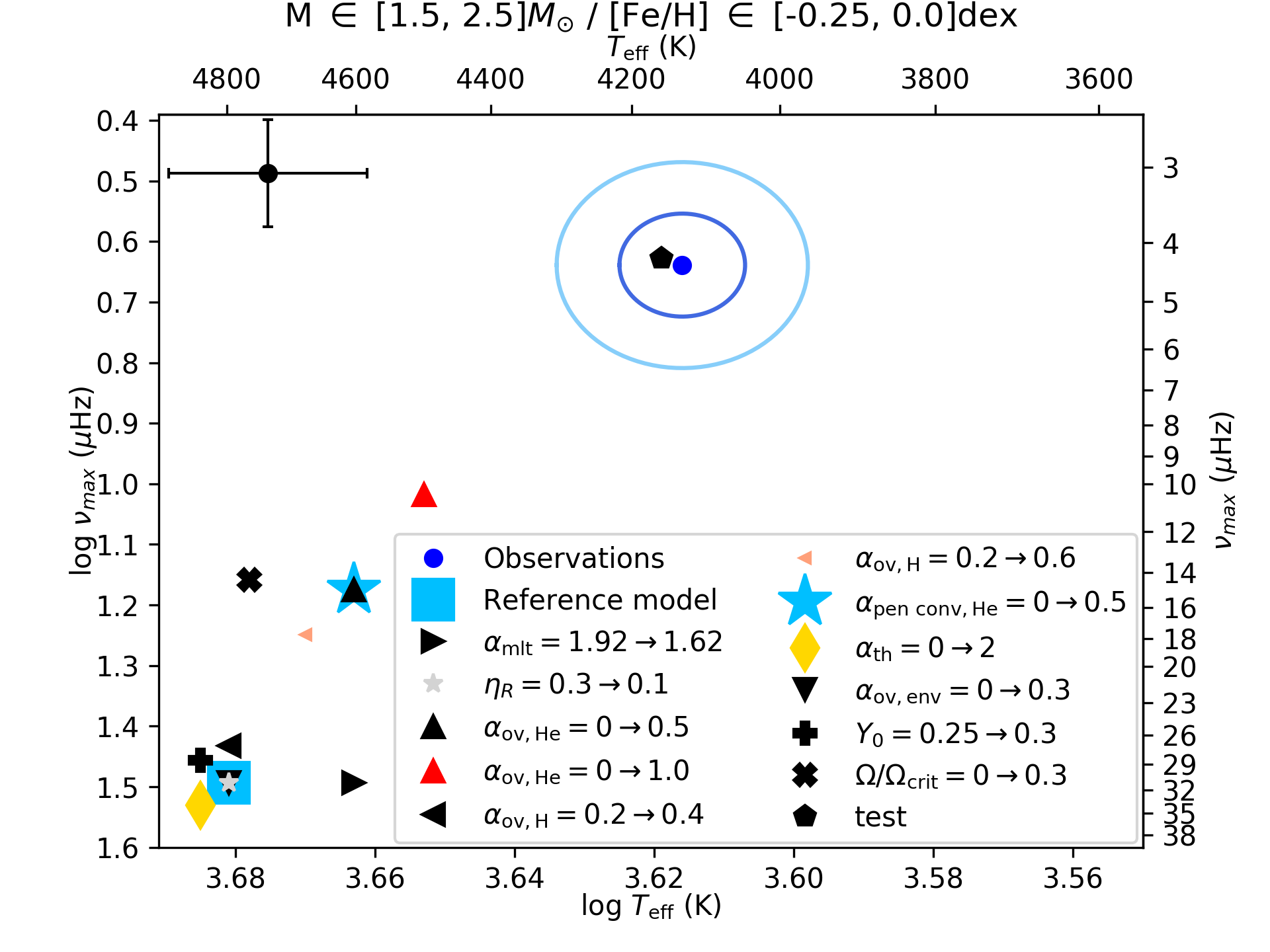}}
		\rotatebox{0}{\includegraphics[width=0.5\linewidth]{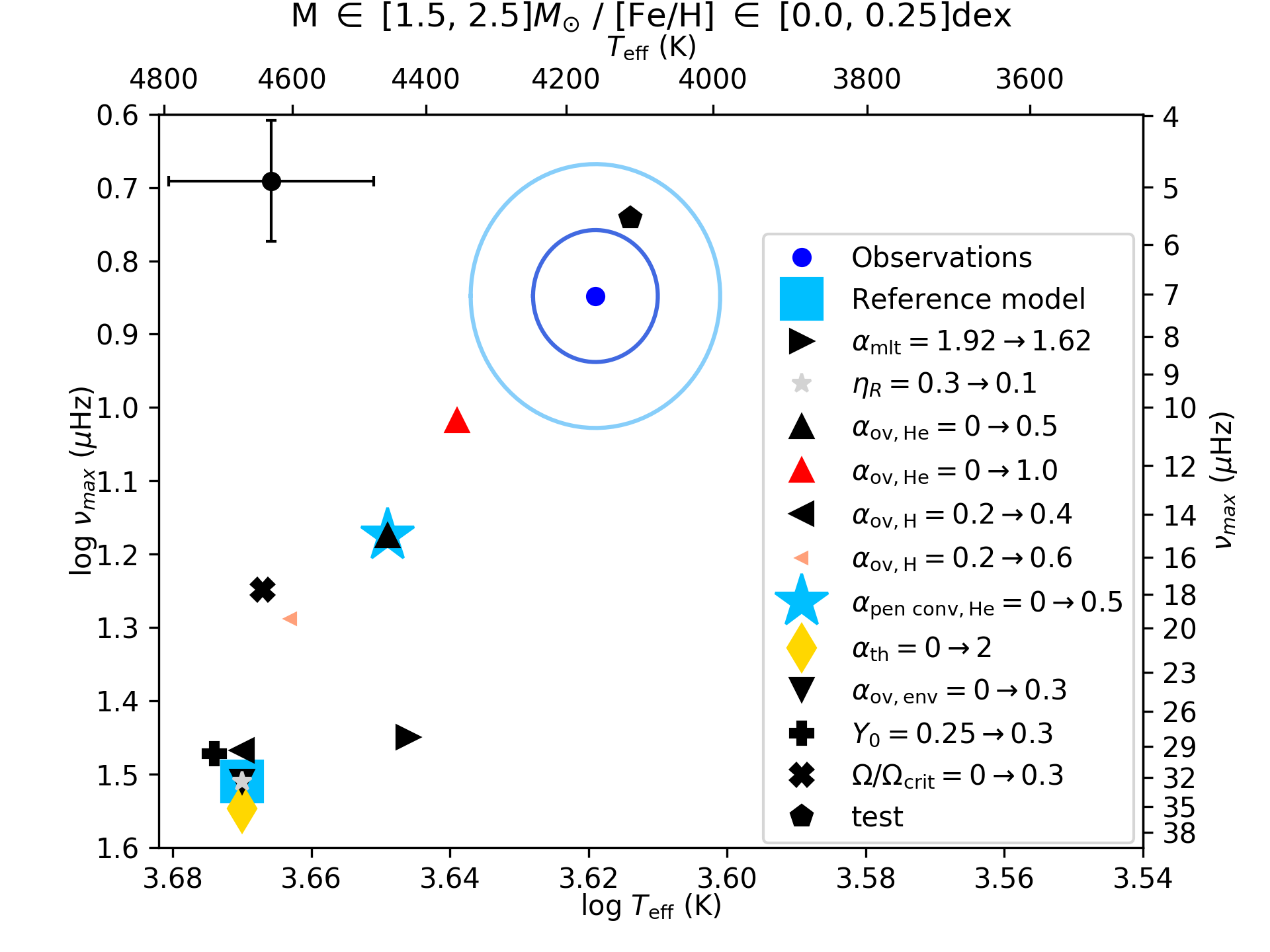}}
	\end{minipage} 
	\small
	\caption{\small{Same label as in Fig.~\ref{fig:loc_AGBb_0_9_to_1_2M}, but for the bins of mass $M \in[1.5,2.5]M_{\odot}$ and metallicity [Fe/H] $\in [-1.0, -0.5]$, $[-0.5, -0.25]$, $[-0.25, 0.0]$, $[0.0, 0.25]\,$dex. An additional model represented by a black cross has been computed to explore the effects of rotation by taking the rotation rate $\Omega_{\mathrm{ZAMS}}/\Omegacrit = 0.3$ during the main sequence, relatively to the reference model, where $\Omegacrit$ is the surface critical angular velocity for the star to be dislocated. Another model labelled `test' and represented by a black pentagon has been computed to check if combining several changes could allow us to reproduce observations. Relatively to the reference model, these changes are the adding of rotation $\Omega_{\mathrm{ZAMS}}/\Omegacrit = 0.3$, He-core overshooting $\alphaovHe = 1.0$, the removal of H-core overshooting $\alphaovH = 0.2 \rightarrow 0$, and the decrease of $\alphaMLT = 1.92 \rightarrow 1.62$. The numerical values are listed in Table~\ref{table:AGBb_location_1_5_2_5M}} in Appendix~\ref{appendix:AGBb_location}. We warn that the range of the axes are not the same between panels.
	}
	\label{fig:loc_AGBb_1_5_to_2_5M}
\end{figure*}

We applied the procedure described in Sect.~\ref{sec:method} both to the data set and the stellar models and examined the AGBb location for all the bins of mass and metallicity previously introduced. The results are presented in Fig.~\ref{fig:loc_AGBb_observations}$-$\ref{fig:loc_AGBb_1_5_to_2_5M} and Tables~\ref{table:AGBb_location_0_6_0_9M}$-$\ref{table:AGBb_location_1_5_2_5M} in Appendix~\ref{appendix:AGBb_location}. \\

\subsection{The AGBb seen from observations}
\label{subsec:AGBb_observations}

From observations, we can highlight a clear mass dependence in the AGBb location: the higher the mass, the lower the $\numax$ associated to the AGBb, whatever the range of metallicity (see left panel of Fig.~\ref{fig:loc_AGBb_observations}). Namely, the higher the mass, the farther the distance between the AGBb and the clump phase along the evolutionary track. We find the AGBb to occur around $\log\numax \sim 0.84$ ($\numax \sim 6.9~\mu$Hz) at $M\sim 1M_{\odot}$ and $\log\numax \sim 0.52$ ($\numax \sim 3.3~\mu$Hz) at $M \sim 2M_{\odot}$, with a typical standard deviation of $\sigma_{2} = 0.06$ and uncertainty on the $\log\numax$ measurements of $\sigma_{\log\numax}$ = 0.02. According to seismic scaling relations, the luminosity $L$ depends on $\numax$ following

\begin{equation}
\label{eq:scaling_relation_nu_max_L}
 \frac{L}{L\ind{\odot}} = \frac{M}{M\ind{\odot}}\left( \frac{\numax}{\nu\ind{max,\odot}} \right)^{-1} \left( \frac{\Teff}{T\ind{eff,\odot}} \right)^{7/2}.
\end{equation}
Then, the higher the mass, the higher the luminosity, which totally agrees with theoretical predictions since the AGBb luminosity has been found to increase with mass at fixed metallicity \citep[][their Fig.~3]{1999ApJ...511..225A}. This behaviour has also been found by Yu et al. 2022 (in prep), who observed that the overdensity of stars associated to the AGBb is shifted between \textit{Kepler}, APOGEE and GALAH stars. This overdensity shift is correlated with a shift of the stellar mass distribution, resulting in a different mean stellar mass in those samples.\\
Besides, we notice in the right panel of Fig.~\ref{fig:loc_AGBb_observations} that the AGBb occurs at lower temperature for high-mass stars (around $\log\Teff \sim 3.630$ at $M\sim 1M_{\odot}$ and $\log\Teff \sim 3.610$ at $M\sim 2M_{\odot}$). Although this temperature dependence is clear in Fig.~\ref{fig:loc_AGBb_observations}, it may be subject to the uncertainty on the $\log\Teff$ measurements of $\sigma_{\log\Teff} = 0.01-0.02$ and our ability to precisely delimit the overdensity, with a standard deviation $\sigma_{2} = 0.01-0.02$. \\

Beyond that mass dependence, we can see in Fig.~\ref{fig:loc_AGBb_observations} a weak metallicity effect on the AGBb location in $\numax$. At fixed mass, the AGBb location in $\numax$ slowly increases with metallicity at low mass ($M \leq 1.2 M_{\odot}$) and noticeably increases at high mass ($M \geq 1.2 M_{\odot}$). This observational trend is consistent with the theoretical results of \cite{1999ApJ...511..225A}. Indeed, for low-mass stars ($M \leq 1.2M_{\odot}$), these authors showed that a change of metallicity does not highly impact the luminosity of the AGBb whereas for high-mass stars ($M \geq 1.2M_{\odot}$) metallicity effects are more important, which is what we observe. Nevertheless, these trends are mainly valid for high-metallicity stars since our sample only contains a small number of metal-poor stars (with [Fe/H] $\leq -0.5\,$dex, see Table~\ref{Table:stars_per_bin}).
Besides, we can see in Fig.~\ref{fig:loc_AGBb_observations} that the AGBb tends to occur at lower $\log\Teff$ when the metallicity increases. However, the $\log\Teff$ variations with metallicity are close to the typical uncertainty on $\log\Teff$, so this behaviour needs to be confirmed. Overall, our results tend to confirm that the AGBb occurrence slightly depends on the metallicity which would make the use of AGBb as standard candle questionable \citep{1992MmSAI..63..485P, 1992MmSAI..63..491F}, at least at high metallicity. This is discussed in Sect.~\ref{sec:discussion}.  \\
To sum up, we find a clear mass dependence of the AGBb location, the higher the mass, the lower $\numax$ and $\Teff$ at which the AGBb occurs, i.e. the later the AGBb occurrence. Moreover, the AGBb tends to occur at slightly higher $\numax$ and lower $\Teff$ for metal-rich stars.

\subsection{The necessity to calibrate the core overshooting parameter}
\label{subsec:calib_low_mass}

In order to estimate the amount of core overshooting $\alphaovHe$ needed to reproduce the observations, we computed evolutionary tracks without core overshooting ($\alphaovHe = 0$). In Fig.~\ref{fig:loc_AGBb_0_6_to_0_9M}$-$\ref{fig:loc_AGBb_1_5_to_2_5M}, we can see that the AGBb locations in observations and models without core overshooting during the clump phase do not overlap at all, neither in the $M-\mathrm{[Fe/H]}-\log\numax$ nor in the $M-\mathrm{[Fe/H]}-\log\Teff$ plane. However, we notice that for a given set of stellar parameters, the larger the mass, the larger the differences between observations and models in $\numax$ and $\Teff$ at the AGBb. This is also what we observe toward low metallicity, but only in $\Teff$. As highlighted in Sect.~\ref{subsec:AGBb_observations} and in theoretical works \citep[][their Fig.~3]{1999ApJ...511..225A}, the metallicity effects are small but still impacts the frequency $\numax$ at which the AGBb occurs. Consequently, we can conclude that models with different masses and metallicities have to be differently calibrated. \\

Then, we adopted a moderate and high core overshooting parameter $\alphaovHe = 0.5$ and $1.0$, which are the values that provide the best matching of the period-spacing and luminosity distributions between observations and stellar models during He-burning phases in the range $M\in [1.3, 1.7]M_{\odot}$ \citep{2015MNRAS.453.2290B}. From Fig.~\ref{fig:loc_AGBb_0_6_to_0_9M}$-$\ref{fig:loc_AGBb_1_5_to_2_5M}, we can conclude that core overshooting has to take place during the core He-burning phase to reproduce observations. Adding core overshooting during the clump phase increases the distance between the latter and the AGBb location along the evolutionary track, which makes the AGBb occur at lower $\numax$ and lower $\Teff$. We note that $\alphaovHe = 0.5$ (respectively $\alphaovHe = 1.0$) gives a nice agreement between models and observations for stellar mass $M\in [0.9, 1.2]M_{\odot}$ (respectively $M\in [1.2, 1.5]M_{\odot}$) for all metallicities. This tends to confirm that the higher the mass, the higher the core overshooting $\alphaovHe$ must be for the models to agree with observations. Nevertheless, for high-mass stars adding He-core overshooting seems inadequate to reproduce observations. The value $\alphaovHe = 1.0$ in units of $H_{P}$, which quantifies the extent of the mixing region beyond the boundary of convective instability, may be unrealistic since the overshooting region then becomes larger than the convective core. Consequently, the inclusion of additional physical processes may be necessary to make models and observations match at high mass. \\
In $\log\Teff$, the AGBb location varies with metallicity at fixed mass in stellar models, see Fig.~\ref{fig:loc_AGBb_0_6_to_0_9M}$-$\ref{fig:loc_AGBb_1_5_to_2_5M} and Tables \ref{table:AGBb_location_0_6_0_9M}$-$\ref{table:AGBb_location_1_5_2_5M}. 
Therefore, other model parameters have to be fine-tuned in order to make observations and models agree both in $\log\numax$ and $\log\Teff$ for all bins of metallicity. This raises the question of degeneracies and uncertainties on stellar parameters. In Sect.~\ref{sec:discussion}, we explore the effects of model input physics that could influence the AGBb location. \\

\section{Discussion}
\label{sec:discussion}

\subsection{Calibration of physical parameters at low mass}

In Sect.~\ref{sec:discussion}, we explore the effects of model input physics that could influence the AGBb. Up to this point, we investigated the impact of He-core overshooting on the AGBb by taking $\nabla_{T} = \gradrad$ in the overshooting region. Following \cite{2015MNRAS.453.2290B}, we also investigated  the penetrative convection scenario defined as $\nabla_{T} = \gradad$ in the overshooting region. According to Fig.~\ref{fig:loc_AGBb_0_6_to_0_9M}$-$\ref{fig:loc_AGBb_1_2_to_1_5M} there is no difference in the location of the AGBb between those two scenarios. \cite{2015MNRAS.453.2290B} reached the same conclusion, nevertheless \cite{2017MNRAS.469.4718B} noticed that the period-spacing distribution of He-burning stars observed by \cite{2016A&A...588A..87V} better matches the one obtained with a radiative transport in the overshooting region. Therefore, seismic constraints support the use of overshooting with radiative transport during core He-burning phase, without any impact on the calibration of the AGBb. \\

\subsubsection{Efficiency of convection}

Besides, by changing the mixing length parameter $\alphaMLT$, we noted that the convection efficiency considerably impacts the AGBb location in $\Teff$. In Fig.~\ref{fig:loc_AGBb_0_6_to_0_9M}$-$\ref{fig:loc_AGBb_1_2_to_1_5M},
we can see that a $\Delta \alphaMLT$ decrease of $0.3$ induces a shift of the AGBb toward low $\Teff$, but marginally modifies its luminosity. In fact, when the mixing length parameter decreases, the energy transport in the envelope is less efficient, the stellar radius $R$ increases and the effective temperature $\Teff$ decreases. Then, the evolutionary track is shifted toward low $\Teff$, including the AGBb location. On the other hand, by considering the scaling relation

\begin{equation}
\label{eq:scaling_relation_nu_max_Teff}
\frac{\numax}{\nu\ind{max,\odot}} \simeq \frac{M}{M\ind{\odot}}\left( \frac{R}{R\ind{\odot}} \right)^{-2} \left( \frac{\Teff}{T\ind{eff,\odot}} \right)^{-1/2},
\end{equation}
we see that at fixed mass, increasing the radius $R$ and decreasing the effective temperature $\Teff$ simultaneously has limited effect on $\numax$, which justifies the minor impact of $\Delta \alphaMLT$ on the AGBb location in $\numax$.

\subsubsection{Other model inputs}

We checked that some physical mechanisms do not modify the AGBb location. They are summarised below:

\begin{itemize}

\item Modifying the H-core overshooting $\alphaovH$ during the MS does not shift the AGBb location for low-mass stars ($M \leq 1.5 M_{\odot}$) since their convective core is either not developed yet or very small.
\item Interestingly, adding an amount of envelope undershooting of $\alphaundersh = 0.3 H_{P}$ (i.e. overshooting from the convective envelope into the radiative core) from the main sequence up to the AGB does not impact the AGBb location, while it does impact the RGBb location \citep{2018ApJ...859..156K}. This implies that the calibrations of mixing processes brought by the RGBb (envelope undershooting) and AGBb (He-core overshooting) are independent. 
\item On the other hand, adding thermohaline convection from the main sequence up to the early AGB with $\alphath = 2$ marginally modifies the AGBb location. Mixing processes between the convective envelope and the radiative core do not seem to significantly impact the AGBb.
\item Changing the mass loss rate on the RGB from $\eta_{R} = 0.3$ to  $\eta_{R} = 0.1$ slightly shifts the AGBb location. This suggests that the changes the star experienced due to mass loss do not impact the AGBb occurrence. Only the final mass reached at the AGBb matters for determining the AGBb location.
\item By varying the initial helium mass fraction from $Y_{0} = 0.253$ to $Y_{0} = 0.303$, the AGBb occurs at slightly lower $\numax$, i.e. at higher luminosity. This is consistent with expectations since an increased initial helium mass fraction enlarges the lifetime of the core He-burning phase. More thermonuclear energy is released, then the luminosity is higher at this evolutionary stage.

\end{itemize}

As a conclusion, we are able to reproduce the AGBb location of low-mass stars with stellar models, particularly by including He-core overshooting $\alphaovHe$ as investigated by \cite{2015MNRAS.453.2290B}. We find that a helium core overshooting parameter $\alphaovHe \in [0.25, 0.50]$ is needed to make observations and models match in the mass bins $M \in [0.6, 0.9], [0.9, 1.2]M_{\odot}$ while $\alphaovHe \in [0.50, 1.0]$ is more appropriate in the mass bin $M \in [1.2, 1.5]M_{\odot}$. Deviations of models from observations in $\Teff$ can be captured by adjusting the mixing length parameter $\alphaMLT$. The main sources of uncertainty on the calibration of He-core  overshooting come from the initial helium mass fraction and potential other mixing processes such as rotational mixing. Additional observational constraints could be used to reduce these uncertainties, in particular the location of the red clump phase. The physical parameter changes we explored also have an impact on the location of the red clump phase, so combining the observed AGBb location with that of the red clump phase would lead to a more precise calibration. In Appendix~\ref{appendix:distance_AGBb_clump}, we explore how physical ingredients impact the ratio of location in $\log \numax$ and $\log \Teff$ between the AGBb and the red clump phase. We note that only the adding of He-core overshooting modifies the distance between the AGBb and the red clump locations along the evolutionary track, while the other parameters leave this distance unchanged. Some parameters not only have an effect on the AGBb location but also on the red clump location, such as the initial helium abundance. Additional work is required to improve the calibration of the simultaneous investigations of the red clump and AGBb locations. On the other hand, we did not explore rotation-induced mixing in low-mass stars since physical ingredients such as surface magnetic braking \citep{2012A&A...537A.146E} are missing to correctly model rotational mixing in low-mass stars ($M \leq 1.5 M_{\odot}$) in the default \mesa\ files.

\subsection{Calibration of physical parameters at high mass}

The location of the AGBb derived from observations and stellar models are represented in Fig.~\ref{fig:loc_AGBb_1_5_to_2_5M} for high-mass stars. Modifying the reference model in a similar way as for low-mass stars leads to the same effects highlighted in Sec.~\ref{subsec:calib_low_mass}. Adding  He-core overshooting increases the distance between the AGBb occurrence and the core He-burning phase along the evolutionary track, decreasing the mixing length parameter $\alphaMLT$ makes the AGBb occur at lower $\Teff$, and modifying the other parameters does not highly impact the AGBb location except for the H-core overshooting parameter. Indeed, for stellar masses above $1.5 M_{\odot}$ the convective core during the main sequence is sufficiently developed so that  H-core overshooting can occur. In Fig.~\ref{fig:loc_AGBb_1_5_to_2_5M}, it can be seen that increasing $\alphaovH$ from $0.2$ to $0.6$ has roughly the same effect on the AGBb location as adding  He-core overshooting $\alphaovHe = 0.5$. 
Nevertheless, a high efficiency of H-core overshooting appears to be unrealistic considering the latest values calibrated with observational constraints in eclipsing binaries \citep[e.g.][]{2016A&A...592A..15C, 2017ApJ...849...18C, 2018ApJ...859..100C, 2019ApJ...876..134C} which do not exceed $\alphaovH \sim 0.2$.  Similarly, values of $\alphaovH$ lower than $0.2$ have been derived from the calibration of dipole modes in low-mass stars by \cite{2016A&A...589A..93D}. Finally, recent theoretical predictions based on 3D numerical hydrodynamics simulations of penetrative convection also give $\alphaovH  < 0.2$ for masses $M < 3\ M_\odot$ \citep{2022ApJ...926..169A, 2022arXiv220309525J}. 
The additional effects of this unrealistic H-core overshooting can be mimicked by taking into account additional mixing processes at work between the convective core and the radiative core. For instance, we found that adding rotational mixing during the main sequence with a rotation rate $\Omega_{\mathrm{ZAMS}}/\Omegacrit = 0.3$ roughly produces the same changes in the AGBb location (see Fig.~\ref{fig:loc_AGBb_1_5_to_2_5M}, for the metallicity bins [Fe/H] $\in [-0.25, 0.0], [0.0, 0.25]\,$dex). \\

None of the physical mechanisms added to the reference model is enough to reproduce observations. Even the  model closest to observations that is obtained by adding a high He-core overshooting cannot reproduce the observed AGBb location. Choosing a higher efficiency of He-core overshooting may be unrealistic since the extent of the extra mixing region would be even higher than that of the convective core, but it suggests that additional mixing processes are needed to match models and observations. To investigate those effects a bit further, we combined the two mixing processes that mostly impact the AGBb location, i.e. core overshooting during the core He-burning phase and rotational mixing during the main sequence, and added them to the reference model. As illustrated in Fig.~\ref{fig:loc_AGBb_1_5_to_2_5M}, by adding rotational mixing with a rotation rate $\Omega_{\mathrm{ZAMS}}/\Omegacrit = 0.3$, He-core overshooting $\alphaovHe = 1.0$, by taking  H-core overshooting away $\alphaovHe = 0$ and decreasing the mixing length parameter of $\Delta \alphaMLT = 0.3$, we are able to reproduce the observed AGBb location. \\
To sum up, several mixing processes such as rotational mixing and He-core overshooting need to be simultaneously taken into account and calibrated to reproduce observations of high-mass stars. However, rotational mixing during the main sequence remains exploratory and further work is required to quantify its significance relatively to other mixing processes such as He-core overshooting.

\subsection{The AGBb as a distance indicator}

In Sect.~\ref{subsec:AGBb_observations}, we noticed that the AGBb location in $\numax$ slightly changes with metallicity at fixed mass, especially for high-mass stars, which implies that the luminosity at the AGBb varies with metallicity. This agrees with the theoretical results of \cite{1999ApJ...511..225A}, where the AGBb luminosity is expected to significantly vary with metallicity, especially for high-mass stars with $M \geq 1.2 M_{\odot}$. At first glance, our conclusions seem to be in disagreement with the results of the models of \cite{1992MmSAI..63..485P} and \cite{1992MmSAI..63..491F} as they justified the use of the AGBb as standard candle by its independence from metallicity. However, these studies are based on a sample of low-metallicity stars in Galactic globular clusters with $\mathrm{[Fe/H]} \lesssim -0.5\,$dex while ours is mainly composed of high-metallicity stars with $\mathrm{[Fe/H]} \gtrsim -0.5\,$dex (see Table~\ref{Table:stars_per_bin}). This tends to confirm that the AGBb location changes at high metallicity. Therefore, metal-rich AGBb stars cannot be used as standard candles. However, this behaviour needs further inspections at low metallicity. We do not have a large enough number of metal-poor AGB stars with [Fe/H] $<-0.75\,$dex to create additional metallicity bins, which limits the analysis of the metallicity dependence of the AGBb.
A larger sample of stars would be desirable to confirm or infirm this metallicity dependence at low metallicity. In parallel, it could be interesting to evaluate the potential of the AGBb as a distance indicator, and test if any metallicity bias is identifiable.

\section{Conclusion}
\label{sec:conclusion}

With the excellent precision of photometric data collected by \Kepler\ and \tess, we are now able to perform asteroseismic studies of high-luminosity red giants. This gives access to oscillation mode properties of those stars, which can be used to constrain stellar interiors. In this work, we took advantage of the $\numax$ estimates from \Kepler\ and \tess\ targets and combined them with spectroscopic data to characterise the AGBb in the widest range and most resolved bins of mass and metallicity explored so far. This would not have been possible without combining targets from several catalogs given the small number of evolved giants subject to a seismic study and the uncertainties on the classification methods between RGB and AGB stars. We detected and accurately located the AGBb in the $\log\Teff - \log\numax$ plane, using a statistical method to distinguish stars belonging to the AGBb and to the AGB background. We highlighted that the occurrence of the AGBb depends on the stellar mass: it clearly takes place at lower $\numax$ (i.e. at higher luminosity) and occurs within uncertainty at cooler temperature for high-mass stars, in agreement with theoretical models. In parallel, the dependence of the AGBb location on metallicity implies that using it as a standard candle requires some care. \\

Then, we were able to use the AGBb location in the $\log\Teff - \log\numax$ plane as a constraint for parameters in stellar models in limited bins of mass and metallicity. Mainly the mixing-length parameter and mixing processes such as He-core overshooting affect the location of the AGBb. Some stellar parameters do not affect the AGBb location, or slightly only, such as the initial helium abundance $Y_{0}$, the mass-loss rate on the RGB $\eta_{R}$, the envelope undershooting $\alphaundersh$ and the thermohaline convection $\alphath$. Those stellar parameters contribute to the uncertainty of the calibration of mixing processes to match observations. We confirmed that models without core overshooting during the core He-burning phase cannot reproduce observations, as already shown in \cite{2015MNRAS.453.2290B}. Moreover, we reported that the amount of He-core overshooting needed to match observations and models depends on the stellar mass, and increases with it. Indicatively, the core overshooting value $\alphaovHe \in [0.25, 0.50]$ nicely suits observations for stars with $M \in [0.6, 1.2]M_{\odot}$ while $\alphaovHe \in [0.5, 1.0]$ better suits those for stars with $M \in [1.2, 1.5]M_{\odot}$. However, for high-mass stars $M \geq 1.5 M_{\odot}$, modifying the  He-core overshooting only does not allow us to reproduce observations. In this case, we explored additional mixing processes, especially rotational mixing during the main sequence, and we found that we could match models and observations by combining rotation-induced mixing and He-core overshooting. Further work is needed to investigate the possible degeneracy between those mixing processes at high mass, and quantify their weight.\\
The core overshooting calibration does also depend on the metallicity, but not as strongly as the mass in terms of $\log\numax$, which makes the values previously cited well suited to all bins of metallicity studied. However, because the AGBb location in $\log\Teff$ varies with metallicity in models, we need to calibrate other parameters such as the mixing-length parameter $\alphaMLT$ in stellar models so that observations and stellar models match in $\log\Teff$ and $\log\numax$ in the same time.\\

In the future, new space-borne missions will be helpful to fill the sample of evolved red giants targets, hence to take into account more low-metallicity stars ([Fe/H]$\leq -0.5\,$dex) and high-mass stars ($M\geq 1.5M_{\odot}$). This will give the opportunity to improve the precision on the observed AGBb location for those bins of mass and metallicity where we lack asteroseismic and spectroscopic data. Besides, it will confirm or disprove the potential of the AGBb to be a suitable standard candle.


\section*{Acknowledgements}

The authors are grateful to the anonymous referee who helped them in improving this paper with constructive suggestions. G.D. thanks S. Khan and O. Hall for their valuable suggestions that helped improving this work. D.B. acknowledges supported by FCT through the research grants UIDB/04434/2020, UIDP/04434/2020 and PTDC/FIS-AST/30389/2017, and by FEDER-Fundo Europeu de Desenvolvimento Regional through COMPETE2020-Programa Operacional Competitividade e Internacionalizaç\~ao (grant: POCI-01-0145-FEDER-030389).


\bibliographystyle{./Outils_latex/aa} 
\bibliography{./Outils_latex/biblio_p} 
\clearpage
\newpage


\newpage

\begin{appendix}

\section{Effects of stellar parameters on the AGBb occurrence}
\label{appendix:AGBb_location}

The AGBb locations in $\log\Teff$ and $\log\numax$ presented in Fig.~\ref{fig:loc_AGBb_0_6_to_0_9M},~\ref{fig:loc_AGBb_0_9_to_1_2M},~\ref{fig:loc_AGBb_1_2_to_1_5M},~\ref{fig:loc_AGBb_1_5_to_2_5M} in the mass bins $M\in [0.6, 0.9]$, $[0.9, 1.2]$, $[1.2, 1.5]$, $[1.5, 2.5]M_{\odot}$ and metallicity bins [Fe/H] $\in [-1.0, -0.5]$, $[-0.5, -0.25]$, $[-0.25, 0.0]$, $[0.0, 0.25]\,$dex are summarised in Table~\ref{table:AGBb_location_0_6_0_9M}, \ref{table:AGBb_location_0_9_1_2M}, \ref{table:AGBb_location_1_2_1_5M}, \ref{table:AGBb_location_1_5_2_5M}. 

\begin{table*}[]
\begin{center}
\caption{AGBb location for observations and models with $M \in [0.6, 0.9]M_{\odot}$}
\begin{tabular}{llllll}
\hline
\hline
$M$ ($M_{\odot}$) & & & $[0.6,0.9]$ & & \\
\hline
[Fe/H] (dex) & &  $[-1.0,-0.5]$ & $[-0.5,-0.25]$ & \ $[-0.25,0.0]$ & \ \ \ $ [0.0,0.25]$ \\
\hline
Observations                                         & $\log \Teff$  & 3.643 $\pm$ 0.019 & 3.639 $\pm$ 0.011 & 3.632 $\pm$ 0.010  & \qquad \ \ - \\
                                                     & $\log \numax$ & 0.831 $\pm$ 0.031 & 0.882 $\pm$ 0.075 & 0.887 $\pm$ 0.077  & \qquad \ \ - \\
\hline
Reference model                                      & $\log \Teff$  & 3.688 $\pm$ 0.011 & 3.674 $\pm$ 0.014 & 3.660 $\pm$ 0.018  & 3.642 $\pm$ 0.014 \\
                                                     & $\log \numax$ & 1.089 $\pm$ 0.060  & 1.089 $\pm$ 0.037 & 1.089 $\pm$ 0.065 & 1.055 $\pm$ 0.064 \\
\hline
$\alphaMLT = 1.92 \rightarrow 1.62$                  & $\log \Teff$  & 3.670 $\pm$ 0.018  & 3.653 $\pm$ 0.019 & 3.635 $\pm$ 0.015 & 3.628 $\pm$ 0.012 \\
                                                     & $\log \numax$ & 1.071 $\pm$ 0.060  & 1.037 $\pm$ 0.076 & 1.003 $\pm$ 0.057 & 1.003 $\pm$ 0.040  \\
\hline
$\eta_{R} = 0.3 \rightarrow 0.1$                     & $\log \Teff$  & 3.681 $\pm$ 0.011 & 3.670 $\pm$ 0.011  & 3.656 $\pm$ 0.015 & 3.642 $\pm$ 0.015 \\
                                                     & $\log \numax$ & 1.123 $\pm$ 0.045 & 1.089 $\pm$ 0.038 & 1.089 $\pm$ 0.069 & 1.089 $\pm$ 0.067 \\
\hline
$\alphaovHe = 0 \rightarrow 0.5$                     & $\log \Teff$  & 3.660 $\pm$ 0.019  & 3.649 $\pm$ 0.022 & 3.632 $\pm$ 0.023 & 3.617 $\pm$ 0.021 \\
                                                     & $\log \numax$ & 0.743 $\pm$ 0.099 & 0.743 $\pm$ 0.099 & 0.711 $\pm$ 0.090  & 0.743 $\pm$ 0.184 \\
\hline
$\alphaovHe = 0 \rightarrow 1.0$                     & $\log \Teff$  & 3.649 $\pm$ 0.020  & 3.632 $\pm$ 0.020  & 3.617 $\pm$ 0.021 & 3.603 $\pm$ 0.018 \\
                                                     & $\log \numax$ & 0.535 $\pm$ 0.108 & 0.504 $\pm$ 0.116 & 0.504 $\pm$ 0.124 & 0.504 $\pm$ 0.120  \\
\hline
$\alphaovH = 0.2 \rightarrow 0.4$                    & $\log \Teff$  & 3.685 $\pm$ 0.014 & 3.678 $\pm$ 0.015 & 3.663 $\pm$ 0.019 & 3.646 $\pm$ 0.017 \\
                                                     & $\log \numax$ & 1.122 $\pm$ 0.053 & 1.087 $\pm$ 0.062 & 1.087 $\pm$ 0.063 & 1.087 $\pm$ 0.044 \\
\hline
$\alphaovH = 0.2 \rightarrow 0.6$                    & $\log \Teff$  & 3.688 $\pm$ 0.013 & 3.678 $\pm$ 0.015 & 3.660 $\pm$ 0.020   & 3.642 $\pm$ 0.016 \\
                                                     & $\log \numax$ & 1.122 $\pm$ 0.050  & 1.087 $\pm$ 0.034 & 1.087 $\pm$ 0.061 & 1.087 $\pm$ 0.071 \\
\hline
$\alpha_{\mathrm{pen\ conv,He}} = 0 \rightarrow 0.5$ & $\log \Teff$  & 3.660 $\pm$ 0.019  & 3.649 $\pm$ 0.022 & 3.632 $\pm$ 0.023 & 3.617 $\pm$ 0.021 \\
                                                     & $\log \numax$ & 0.743 $\pm$ 0.099 & 0.743 $\pm$ 0.099 & 0.711 $\pm$ 0.090  & 0.743 $\pm$ 0.184 \\
\hline
$\alphath = 0 \rightarrow 2$                         & $\log \Teff$  & 3.688 $\pm$ 0.013 & 3.674 $\pm$ 0.014 & 3.660 $\pm$ 0.013  & 3.646 $\pm$ 0.014 \\
                                                     & $\log \numax$ & 1.157 $\pm$ 0.062 & 1.123 $\pm$ 0.073 & 1.123 $\pm$ 0.072 & 1.089 $\pm$ 0.072 \\
\hline
$\alphaundersh = 0 \rightarrow 0.3$                  & $\log \Teff$  & 3.685 $\pm$ 0.014 & 3.674 $\pm$ 0.014 & 3.660 $\pm$ 0.017  & 3.642 $\pm$ 0.015 \\
                                                     & $\log \numax$ & 1.089 $\pm$ 0.065 & 1.089 $\pm$ 0.044 & 1.089 $\pm$ 0.040  & 1.055 $\pm$ 0.080  \\
\hline
$Y_{0} = 0.253 \rightarrow 0.303$                       & $\log \Teff$  & 3.685 $\pm$ 0.014 & 3.670 $\pm$ 0.012  & 3.660 $\pm$ 0.013  & 3.642 $\pm$ 0.015 \\
                                                     & $\log \numax$ & 1.030 $\pm$ 0.061  & 1.030 $\pm$ 0.062  & 0.997 $\pm$ 0.059 & 0.997 $\pm$ 0.077 \\
\hline
\label{table:AGBb_location_0_6_0_9M}
\end{tabular}
\\
\end{center}
\textbf{Notes:} The AGBb locations are plotted in Fig.~\ref{fig:loc_AGBb_0_6_to_0_9M}. Models have been obtained by individually changing the parameters of the reference model. These changes are indicated by the arrow.
\end{table*}

\begin{table*}[]
\begin{center}
\caption{AGBb location for observations and models with $M \in [0.9, 1.2]M_{\odot}$}
\begin{tabular}{llllll}
\hline
\hline
$M$ ($M_{\odot}$) & & & $[0.9,1.2]$ & & \\
\hline
[Fe/H] (dex) & &  $[-1.0,-0.5]$ & $[-0.5,-0.25]$ & \ $[-0.25,0.0]$ & \ \ \ $ [0.0,0.25]$ \\
\hline
Observations                                         & $\log \Teff$  & 3.645 $\pm$ 0.015 & 3.636 $\pm$ 0.010 & 3.629 $\pm$ 0.010  & 3.624 $\pm$ 0.008 \\
                                                     & $\log \numax$ & 0.783 $\pm$ 0.054 & 0.860 $\pm$ 0.078 & 0.869 $\pm$ 0.086 & 0.924 $\pm$ 0.076 \\
\hline
Reference model                                      & $\log \Teff$  & 3.681 $\pm$ 0.010 & 3.670 $\pm$ 0.011 & 3.656 $\pm$ 0.013 & 3.642 $\pm$ 0.014 \\
                                                     & $\log \numax$ & 1.190 $\pm$ 0.048 & 1.157 $\pm$ 0.066 & 1.157 $\pm$ 0.071 & 1.157 $\pm$ 0.063 \\
\hline
$\alphaMLT = 1.92 \rightarrow 1.62$                  & $\log \Teff$  & 3.667 $\pm$ 0.016 & 3.649 $\pm$ 0.013 & 3.635 $\pm$ 0.014 & 3.621 $\pm$ 0.013 \\
                                                     & $\log \numax$ & 1.138 $\pm$ 0.062 & 1.104 $\pm$ 0.058 & 1.104 $\pm$ 0.051 & 1.071 $\pm$ 0.048 \\
\hline
$\eta_{R} = 0.3 \rightarrow 0.1$                     & $\log \Teff$  & 3.681 $\pm$ 0.010 & 3.670 $\pm$ 0.010 & 3.656 $\pm$ 0.013 & 3.642 $\pm$ 0.013 \\
                                                     & $\log \numax$ & 1.190 $\pm$ 0.050 & 1.190 $\pm$ 0.057 & 1.157 $\pm$ 0.070 & 1.157 $\pm$ 0.063 \\
\hline
$\alphaovHe = 0 \rightarrow 0.5$                     & $\log \Teff$  & 3.663 $\pm$ 0.016 & 3.649 $\pm$ 0.017 & 3.632 $\pm$ 0.019 & 3.617 $\pm$ 0.019 \\
                                                     & $\log \numax$ & 0.871 $\pm$ 0.055 & 0.839 $\pm$ 0.042 & 0.807 $\pm$ 0.062 & 0.807 $\pm$ 0.082 \\
\hline
$\alphaovHe = 0 \rightarrow 1.0$                     & $\log \Teff$  & 3.653 $\pm$ 0.018 & 3.635 $\pm$ 0.019 & 3.621 $\pm$ 0.018 & 3.603 $\pm$ 0.018 \\
                                                     & $\log \numax$ & 0.659 $\pm$ 0.112 & 0.628 $\pm$ 0.114 & 0.597 $\pm$ 0.110 & 0.597 $\pm$ 0.110 \\
\hline
$\alphaovH = 0.2 \rightarrow 0.4$                    & $\log \Teff$  & 3.685 $\pm$ 0.011 & 3.670 $\pm$ 0.011 & 3.660 $\pm$ 0.013 & 3.639 $\pm$ 0.014 \\
                                                     & $\log \numax$ & 1.225 $\pm$ 0.076 & 1.191 $\pm$ 0.073 & 1.156 $\pm$ 0.065 & 1.156 $\pm$ 0.068\\
\hline
$\alphaovH = 0.2 \rightarrow 0.6$                    & $\log \Teff$  & 3.685 $\pm$ 0.009 & 3.674 $\pm$ 0.012 & 3.660 $\pm$ 0.013 & 3.646 $\pm$ 0.015 \\
                                                     & $\log \numax$ & 1.260 $\pm$ 0.058 & 1.225 $\pm$ 0.071 & 1.191 $\pm$ 0.073 & 1.191 $\pm$ 0.084\\
\hline
$\alpha_{\mathrm{pen\ conv,He}} = 0 \rightarrow 0.5$ & $\log \Teff$  & 3.663 $\pm$ 0.016 & 3.649 $\pm$ 0.017 & 3.632 $\pm$ 0.019 & 3.617 $\pm$ 0.019 \\
                                                     & $\log \numax$ & 0.871 $\pm$ 0.055 & 0.839 $\pm$ 0.042 & 0.807 $\pm$ 0.062 & 0.807 $\pm$ 0.082\\
\hline
$\alphath = 0 \rightarrow 2$                         & $\log \Teff$  & 3.685 $\pm$ 0.009 & 3.670 $\pm$ 0.011 & 3.660 $\pm$ 0.012 & 3.642 $\pm$ 0.014\\
                                                     & $\log \numax$ & 1.190 $\pm$ 0.065 & 1.190 $\pm$ 0.068 & 1.190 $\pm$ 0.065 & 1.157 $\pm$ 0.069 \\
\hline
$\alphaundersh = 0 \rightarrow 0.3$                  & $\log \Teff$  & 3.685 $\pm$ 0.010 & 3.670 $\pm$ 0.011 & 3.656 $\pm$ 0.013 & 3.642 $\pm$ 0.013 \\
                                                     & $\log \numax$ & 1.190 $\pm$ 0.057 & 1.157 $\pm$ 0.043 & 1.157 $\pm$ 0.043 & 1.157 $\pm$ 0.073\\
\hline
$Y_{0} = 0.253 \rightarrow 0.3$                       & $\log \Teff$  & 3.685 $\pm$ 0.009 & 3.674 $\pm$ 0.010 & 3.660 $\pm$ 0.013 & 3.642 $\pm$ 0.013 \\
                                                     & $\log \numax$ & 1.097 $\pm$ 0.066 & 1.097 $\pm$ 0.067 & 1.063 $\pm$ 0.073 & 1.063 $\pm$ 0.077 \\
\hline
\label{table:AGBb_location_0_9_1_2M}
\end{tabular}
\\
\end{center}
\textbf{Notes:} The AGBb locations are plotted in Fig.~\ref{fig:loc_AGBb_0_9_to_1_2M}. Models have been obtained by individually changing the parameters of the reference model. These changes are indicated in the first column.
\end{table*}

\begin{table*}[]
\begin{center}
\caption{AGBb location for observations and models with $M \in [1.2, 1.5]M_{\odot}$}
\begin{tabular}{llllll}
\hline
\hline
$M$ ($M_{\odot}$) & & & $[1.2,1.5]$ & & \\
\hline
[Fe/H] (dex) & &  $[-1.0,-0.5]$ & $[-0.5,-0.25]$ & \ $[-0.25,0.0]$ & \ \ \ $ [0.0,0.25]$ \\
\hline
Observations                                         & $\log \Teff$  & 3.638 $\pm$ 0.015 & 3.637 $\pm$ 0.012 & 3.621 $\pm$ 0.009 & 3.622 $\pm$ 0.009 \\
                                                     & $\log \numax$ & 0.674 $\pm$ 0.046 & 0.781 $\pm$ 0.091 & 0.712 $\pm$ 0.058 & 0.935 $\pm$ 0.084 \\
\hline
Reference model                                      & $\log \Teff$  & 3.688 $\pm$ 0.011 & 3.678 $\pm$ 0.010 & 3.667 $\pm$ 0.012 & 3.653 $\pm$ 0.011 \\
                                                     & $\log \numax$ & 1.284 $\pm$ 0.033 & 1.284 $\pm$ 0.078 & 1.284 $\pm$ 0.078 & 1.319 $\pm$ 0.074 \\
\hline
$\alphaMLT = 1.92 \rightarrow 1.62$                  & $\log \Teff$  & 3.681 $\pm$ 0.014 & 3.660 $\pm$ 0.013 & 3.646 $\pm$ 0.014 & 3.632 $\pm$ 0.016 \\
                                                     & $\log \numax$ & 1.288 $\pm$ 0.066 & 1.288 $\pm$ 0.066 & 1.252 $\pm$ 0.060 & 1.252 $\pm$ 0.063 \\
\hline
$\eta_{R} = 0.3 \rightarrow 0.1$                     & $\log \Teff$  & 3.688 $\pm$ 0.010 & 3.678 $\pm$ 0.011 & 3.663 $\pm$ 0.014 & 3.649 $\pm$ 0.013 \\
                                                     & $\log \numax$ & 1.284 $\pm$ 0.066 & 1.284 $\pm$ 0.072 & 1.249 $\pm$ 0.079 & 1.249 $\pm$ 0.071 \\
\hline
$\alphaovHe = 0 \rightarrow 0.5$                     & $\log \Teff$  & 3.667 $\pm$ 0.015 & 3.656 $\pm$ 0.011 & 3.642 $\pm$ 0.019 & 3.624 $\pm$ 0.019 \\
                                                     & $\log \numax$ & 0.935 $\pm$ 0.078 & 0.935 $\pm$ 0.081 & 1.000 $\pm$ 0.071 & 0.935 $\pm$ 0.057 \\
\hline
$\alphaovHe = 0 \rightarrow 1.0$                     & $\log \Teff$  & 3.656 $\pm$ 0.016 & 3.646 $\pm$ 0.015 & 3.632 $\pm$ 0.019 & 3.614 $\pm$ 0.017 \\
                                                     & $\log \numax$ & 0.753 $\pm$ 0.108 & 0.721 $\pm$ 0.112 & 0.753 $\pm$ 0.089 & 0.753 $\pm$ 0.092 \\
\hline
$\alphaovH = 0.2 \rightarrow 0.4$                    & $\log \Teff$  & 3.688 $\pm$ 0.010 & 3.678 $\pm$ 0.011 & 3.670 $\pm$ 0.012 & 3.653 $\pm$ 0.013 \\
                                                     & $\log \numax$ & 1.323 $\pm$ 0.051 & 1.288 $\pm$ 0.053 & 1.323 $\pm$ 0.052 & 1.323 $\pm$ 0.095 \\
\hline
$\alphaovH = 0.2 \rightarrow 0.6$                    & $\log \Teff$  & 3.688 $\pm$ 0.009 & 3.678 $\pm$ 0.010 & 3.667 $\pm$ 0.013 & 3.653 $\pm$ 0.014 \\
                                                     & $\log \numax$ & 1.323 $\pm$ 0.074 & 1.323 $\pm$ 0.088 & 1.323 $\pm$ 0.097 & 1.288 $\pm$ 0.089 \\
\hline
$\alpha_{\mathrm{pen\ conv,He}} = 0 \rightarrow 0.5$ & $\log \Teff$  & 3.667 $\pm$ 0.015 & 3.656 $\pm$ 0.011 & 3.642 $\pm$ 0.019 & 3.624 $\pm$ 0.019 \\
                                                     & $\log \numax$ & 0.935 $\pm$ 0.078 & 0.935 $\pm$ 0.081 & 1.000 $\pm$ 0.071 & 0.935 $\pm$ 0.057 \\
\hline
$\alphath = 0 \rightarrow 2$                         & $\log \Teff$  & 3.688 $\pm$ 0.010 & 3.678 $\pm$ 0.010 & 3.667 $\pm$ 0.010 & 3.653 $\pm$ 0.014 \\
                                                     & $\log \numax$ & 1.284 $\pm$ 0.023 & 1.284 $\pm$ 0.060 & 1.319 $\pm$ 0.076 & 1.319 $\pm$ 0.083 \\
\hline
$\alphaundersh = 0 \rightarrow 0.3$                  & $\log \Teff$  & 3.688 $\pm$ 0.011 & 3.678 $\pm$ 0.010 & 3.667 $\pm$ 0.011 & 3.653 $\pm$ 0.015 \\
                                                     & $\log \numax$ & 1.284 $\pm$ 0.066 & 1.284 $\pm$ 0.063 & 1.319 $\pm$ 0.050 & 1.284 $\pm$ 0.049 \\
\hline
$Y_{0} = 0.253 \rightarrow 0.303$                       & $\log \Teff$  & 3.688 $\pm$ 0.010 & 3.678 $\pm$ 0.012 & 3.667 $\pm$ 0.014 & 3.656 $\pm$ 0.011 \\
                                                     & $\log \numax$ & 1.252 $\pm$ 0.105 & 1.217 $\pm$ 0.133 & 1.183 $\pm$ 0.073 & 1.183 $\pm$ 0.095 \\
\hline
\label{table:AGBb_location_1_2_1_5M}
\end{tabular}
\\
\end{center}
\textbf{Notes:} The AGBb locations are plotted in Fig.~\ref{fig:loc_AGBb_1_2_to_1_5M}. Models have been obtained by individually changing the parameters of the reference model. These changes are indicated in the first column.
\end{table*}

\begin{table*}[]
\begin{center}
\caption{AGBb location for observations and models with $M \in [1.5, 2.5]M_{\odot}$}
\begin{tabular}{llllll}
\hline
\hline
$M$ ($M_{\odot}$) & & & $[1.5,2.5]$ & & \\
\hline
[Fe/H] (dex) & &  $[-1.0,-0.5]$ & $[-0.5,-0.25]$ & \ $[-0.25,0.0]$ & \ \ \ $ [0.0,0.25]$ \\
\hline
Observations                                         & $\log \Teff$  & 3.624 $\pm$ 0.010 & 3.632 $\pm$ 0.013 & 3.616 $\pm$ 0.009 & 3.619 $\pm$ 0.009 \\
                                                     & $\log \numax$ & 0.609 $\pm$ 0.064 & 0.677 $\pm$ 0.090 & 0.639 $\pm$ 0.085 & 0.848 $\pm$ 0.090 \\
\hline
Reference model                                      & $\log \Teff$  & 3.702 $\pm$ 0.009 & 3.692 $\pm$ 0.009 & 3.681 $\pm$ 0.013 & 3.670 $\pm$ 0.014 \\
                                                     & $\log \numax$ & 1.493 $\pm$ 0.065 & 1.531 $\pm$ 0.065 & 1.493 $\pm$ 0.062 & 1.510 $\pm$ 0.088 \\
\hline
$\alphaMLT = 1.92 \rightarrow 1.62$                  & $\log \Teff$  & 3.685 $\pm$ 0.012 & 3.674 $\pm$ 0.012 & 3.663 $\pm$ 0.013 & 3.646 $\pm$ 0.015 \\
                                                     & $\log \numax$ & 1.450 $\pm$ 0.037 & 1.450 $\pm$ 0.078 & 1.493 $\pm$ 0.089 & 1.450 $\pm$ 0.071 \\
\hline
$\eta_{R} = 0.3 \rightarrow 0.1$                     & $\log \Teff$  & 3.702 $\pm$ 0.009 & 3.692 $\pm$ 0.009 & 3.681 $\pm$ 0.012 & 3.670 $\pm$ 0.014 \\
                                                     & $\log \numax$ & 1.493 $\pm$ 0.027 & 1.493 $\pm$ 0.086 & 1.493 $\pm$ 0.085 & 1.510 $\pm$ 0.080 \\
\hline
$\alphaovHe = 0 \rightarrow 0.5$                     & $\log \Teff$  & 3.681 $\pm$ 0.011 & 3.674 $\pm$ 0.011 & 3.663 $\pm$ 0.015 & 3.649 $\pm$ 0.016 \\
                                                     & $\log \numax$ & 1.175 $\pm$ 0.042 & 1.175 $\pm$ 0.062 & 1.175 $\pm$ 0.070 & 1.175 $\pm$ 0.075 \\
\hline
$\alphaovHe = 0 \rightarrow 1.0$                     & $\log \Teff$  & 3.674 $\pm$ 0.013 & 3.663 $\pm$ 0.013 & 3.653 $\pm$ 0.017 & 3.639 $\pm$ 0.017 \\
                                                     & $\log \numax$ & 1.084 $\pm$ 0.082 & 1.051 $\pm$ 0.073 & 1.018 $\pm$ 0.084 & 1.018 $\pm$ 0.112 \\
\hline
$\alphaovH = 0.2 \rightarrow 0.4$                    & $\log \Teff$  & 3.695 $\pm$ 0.010 & 3.688 $\pm$ 0.010 & 3.681 $\pm$ 0.012 & 3.670 $\pm$ 0.014 \\
                                                     & $\log \numax$ & 1.355 $\pm$ 0.067 & 1.392 $\pm$ 0.087 & 1.432 $\pm$ 0.069 & 1.468 $\pm$ 0.095 \\
\hline
$\alphaovH = 0.2 \rightarrow 0.6$                    & $\log \Teff$  & 3.685 $\pm$ 0.013 & 3.681 $\pm$ 0.012 & 3.670 $\pm$ 0.011 & 3.663 $\pm$ 0.012 \\
                                                     & $\log \numax$ & 1.116 $\pm$ 0.155 & 1.221 $\pm$ 0.110 & 1.249 $\pm$ 0.060 & 1.288 $\pm$ 0.088 \\
\hline
$\alpha_{\mathrm{pen\ conv,He}} = 0 \rightarrow 0.5$ & $\log \Teff$  & 3.681 $\pm$ 0.011 & 3.674 $\pm$ 0.011 & 3.663 $\pm$ 0.015 & 3.649 $\pm$ 0.016 \\
                                                     & $\log \numax$ & 1.175 $\pm$ 0.042 & 1.175 $\pm$ 0.062 & 1.175 $\pm$ 0.070 & 1.175 $\pm$ 0.075 \\
\hline
$\alphath = 0 \rightarrow 2$                         & $\log \Teff$  & 3.702 $\pm$ 0.008 & 3.695 $\pm$ 0.007 & 3.685 $\pm$ 0.011 & 3.670 $\pm$ 0.014 \\
                                                     & $\log \numax$ & 1.493 $\pm$ 0.090 & 1.531 $\pm$ 0.080 & 1.531 $\pm$ 0.082 & 1.547 $\pm$ 0.090 \\
\hline
$\alphaundersh = 0 \rightarrow 0.3$                  & $\log \Teff$  & 3.702 $\pm$ 0.010 & 3.692 $\pm$ 0.010 & 3.681 $\pm$ 0.011 & 3.670 $\pm$ 0.014 \\
                                                     & $\log \numax$ & 1.493 $\pm$ 0.079 & 1.531 $\pm$ 0.077 & 1.493 $\pm$ 0.085 & 1.510 $\pm$ 0.057 \\
\hline
$Y_{0} = 0.253 \rightarrow 0.303$                       & $\log \Teff$  & 3.706 $\pm$ 0.013 & 3.695 $\pm$ 0.012 & 3.685 $\pm$ 0.014 & 3.674 $\pm$ 0.014 \\
                                                     & $\log \numax$ & 1.381 $\pm$ 0.129 & 1.456 $\pm$ 0.102 & 1.456 $\pm$ 0.108 & 1.472 $\pm$ 0.086 \\
\hline
$\Omega_{\mathrm{ZAMS}}/\Omegacrit = 0 \rightarrow 0.3$                       & $\log \Teff$  & 3.688 $\pm$ 0.006 & 3.678 $\pm$ 0.009 & 3.678 $\pm$ 0.011 & 3.667 $\pm$ 0.013 \\
                                                 & $\log \numax$ & 0.833 $\pm$ 0.061 & 0.894 $\pm$ 0.017 & 1.159 $\pm$ 0.091 & 1.249 $\pm$ 0.081 \\
\hline
test                      & $\log \Teff$ & 3.619 $\pm$ 0.017 & 3.622 $\pm$ 0.035 & 3.619 $\pm$ 0.030 & 3.614 $\pm$ 0.019 \\
                                                & $\log \numax$  & 0.559 $\pm$ 0.066 & 0.559 $\pm$ 0.059 & 0.628 $\pm$ 0.167 & 0.742 $\pm$ 0.077 \\
\hline
\label{table:AGBb_location_1_5_2_5M}
\end{tabular}
\\
\end{center}
\textbf{Notes:} The AGBb locations are plotted in Fig.~\ref{fig:loc_AGBb_1_5_to_2_5M}. Models have been obtained by individually changing the parameters of the reference model. These changes are indicated in the first column. The model labelled `test' is obtained with the following changes: $\Omega_{\mathrm{ZAMS}}/\Omegacrit = 0 \rightarrow 0.3$, $\alphaovH = 0.2 \rightarrow 0$, $\alphaovHe = 0 \rightarrow 1.0$, and $\alphaMLT = 1.92 \rightarrow 1.62$.
\end{table*}

\section{Distance between the AGBb and clump phase}
\label{appendix:distance_AGBb_clump}

Another relevant property of the AGBb to investigate is the distance in $\log \numax$ and $\log \Teff$ between its location and that of the core He-burning phase. Indeed, theoretical models report a weak dependence of the luminosity ratio between the AGBb and red clump locations on the metallicity and initial helium abundance \citep{1991ApJS...76..911C, 1995A&A...297..115B}. Accordingly, we inspected how the ratio of $\log \numax$ and $\log\Teff$ between the AGBb and red clump phase varies with a change in physical parameters. To this end, we needed to extract the location of the red clump. We proceeded in the same way as that described in Sect.~\ref{sec:method_subsec_models}, but we adapted it for the overdensity of stars in the clump phase. We only included stellar models for which the core helium abundance lies in the interval $Y_{\mathrm{c}} \in [0.01, 0.95]$ in the histogram. Then, we extracted the clump location independently from that of the AGBb. The ratios in $\log \numax$ and $\log \Teff$ between the AGBb and the red clump phase for our set of stellar models are shown in Table~\ref{table:ratio_AGBb_clump_location_0_6_0_9M}, \ref{table:ratio_AGBb_clump_location_0_9_1_2M}, \ref{table:ratio_AGBb_clump_location_1_2_1_5M}, \ref{table:ratio_AGBb_clump_location_1_5_2_5M}. \\

Overall, the ratio in $\log \Teff$ is almost constant and the ratio in $\log \numax$ weakly decreases (equivalently the ratio in $\log L$ weakly increases) when the metallicity increases with a given set of physical ingredients. This agrees with the typical difference between metal-poor and metal-rich models obtained with the theoretical models of \cite{1991ApJS...76..911C} (their Fig. 7). These ratios do not significantly change within uncertainties when a specific change in physical parameter is performed, except when adding He-core overshooting. Indeed, both ratios in $\log \numax$ and $\log \Teff$ substantially decrease (namely the ratio in $\log L$ increases) when adding He-core overshooting. This implies that the adding of He-core overshooting causes an increase of the distance between the AGBb and the red clump locations along the evolutionary track, but a change in other physical parameters leave this distance constant. \\
Given that some physical ingredients have an effect on the AGBb location but leave the ratio of location between the AGBb and the red clump unchanged, it means that some of those physical ingredients also impact the red clump location. This is not surprising for physical parameters such as the initial helium abundance, as it determines how much helium burning contributes to the stellar luminosity during the red clump phase. Consequently, this ratio in $\log \numax$ could also be used in combination with the AGBb location in $\log \numax$ as calibrators for mixing processes to reproduce both the AGBb and the red clump locations in the same time.

\begin{table*}[]
\begin{center}
\caption{Ratio between the AGBb location and the clump location for models with $M \in [0.6, 0.9]M_{\odot}$}
\begin{tabular}{llllll}
\hline
\hline
$M$ ($M_{\odot}$) & & & $[0.6,0.9]$ & & \\
\hline
[Fe/H] (dex) & &  $[-1.0,-0.5]$ & $[-0.5,-0.25]$ & \ $[-0.25,0.0]$ & \ \ \ $ [0.0,0.25]$ \\
\hline
Reference model                                      & $\ratioTeffAGBbclump$  & 0.993 $\pm$ 0.006 & 0.993 $\pm$ 0.008 & 0.993 $\pm$ 0.008 & 0.992 $\pm$ 0.007 \\
                                                     & $\rationumaxAGBbclump$ & 0.757 $\pm$ 0.058 & 0.757 $\pm$ 0.042 & 0.757 $\pm$ 0.061 & 0.734 $\pm$ 0.060 \\
                                                     \hline
$\alphaMLT = 1.92 \rightarrow 1.62$                  & $\ratioTeffAGBbclump$  & 0.994 $\pm$ 0.009 & 0.993 $\pm$ 0.008 & 0.993 $\pm$ 0.007 & 0.993 $\pm$ 0.005 \\
                                                     & $\rationumaxAGBbclump$ & 0.767 $\pm$ 0.061 & 0.743 $\pm$ 0.074 & 0.741 $\pm$ 0.057 & 0.741 $\pm$ 0.043 \\
                                                     \hline
$\eta_{R} = 0.3 \rightarrow 0.1$                     & $\ratioTeffAGBbclump$  & 0.994 $\pm$ 0.005 & 0.994 $\pm$ 0.005 & 0.993 $\pm$ 0.007 & 0.993 $\pm$ 0.007 \\
                                                     & $\rationumaxAGBbclump$ & 0.781 $\pm$ 0.041 & 0.757 $\pm$ 0.036 & 0.757 $\pm$ 0.059 & 0.757 $\pm$ 0.061 \\
                                                     \hline
$\alphaovHe = 0 \rightarrow 0.5$                     & $\ratioTeffAGBbclump$  & 0.986 $\pm$ 0.009 & 0.986 $\pm$ 0.009 & 0.985 $\pm$ 0.009 & 0.985 $\pm$ 0.008 \\
                                                     & $\rationumaxAGBbclump$ & 0.502 $\pm$ 0.076 & 0.502 $\pm$ 0.080 & 0.480 $\pm$ 0.073 & 0.502 $\pm$ 0.133 \\
                                                     \hline
$\alphaovHe = 0 \rightarrow 1.0$                     & $\ratioTeffAGBbclump$  & 0.985 $\pm$ 0.007 & 0.982 $\pm$ 0.007 & 0.980 $\pm$ 0.008 & 0.981 $\pm$ 0.008 \\
                                                     & $\rationumaxAGBbclump$ & 0.372 $\pm$ 0.081 & 0.350 $\pm$ 0.086 & 0.350 $\pm$ 0.093 & 0.350 $\pm$ 0.091 \\
                                                     \hline
$\alphaovH = 0.2 \rightarrow 0.4$                    & $\ratioTeffAGBbclump$  & 0.994 $\pm$ 0.007 & 0.995 $\pm$ 0.007 & 0.994 $\pm$ 0.008 & 0.994 $\pm$ 0.007 \\
                                                     & $\rationumaxAGBbclump$ & 0.780 $\pm$ 0.054 & 0.756 $\pm$ 0.058 & 0.756 $\pm$ 0.057 & 0.756 $\pm$ 0.045 \\
                                                     \hline
$\alphaovH = 0.2 \rightarrow 0.6$                    & $\ratioTeffAGBbclump$  & 0.994 $\pm$ 0.006 & 0.994 $\pm$ 0.007 & 0.993 $\pm$ 0.006 & 0.993 $\pm$ 0.007 \\
                                                     & $\rationumaxAGBbclump$ & 0.780 $\pm$ 0.049 & 0.756 $\pm$ 0.038 & 0.756 $\pm$ 0.058 & 0.756 $\pm$ 0.066 \\
                                                     \hline
$\alpha_{\mathrm{pen\ conv,He}} = 0 \rightarrow 0.5$ & $\ratioTeffAGBbclump$  & 0.986 $\pm$ 0.009 & 0.986 $\pm$ 0.009 & 0.985 $\pm$ 0.009 & 0.985 $\pm$ 0.008 \\
                                                     & $\rationumaxAGBbclump$ & 0.502 $\pm$ 0.076 & 0.502 $\pm$ 0.080 & 0.480 $\pm$ 0.073 & 0.502 $\pm$ 0.133 \\
                                                     \hline
$\alphath = 0 \rightarrow 2$                         & $\ratioTeffAGBbclump$  & 0.995 $\pm$ 0.007 & 0.995 $\pm$ 0.006 & 0.994 $\pm$ 0.006 & 0.994 $\pm$ 0.007 \\
                                                     & $\rationumaxAGBbclump$ & 0.805 $\pm$ 0.059 & 0.781 $\pm$ 0.064 & 0.781 $\pm$ 0.063 & 0.757 $\pm$ 0.064 \\
                                                     \hline
$\alphaundersh = 0 \rightarrow 0.3$                  & $\ratioTeffAGBbclump$  & 0.994 $\pm$ 0.007 & 0.994 $\pm$ 0.007 & 0.993 $\pm$ 0.007 & 0.993 $\pm$ 0.007 \\
                                                     & $\rationumaxAGBbclump$ & 0.757 $\pm$ 0.060 & 0.757 $\pm$ 0.045 & 0.757 $\pm$ 0.043 & 0.734 $\pm$ 0.068 \\
                                                     \hline
$Y_{0} = 0.253 \rightarrow 0.303$                    & $\ratioTeffAGBbclump$  & 0.995 $\pm$ 0.006 & 0.994 $\pm$ 0.005 & 0.994 $\pm$ 0.006 & 0.993 $\pm$ 0.007 \\
                                                     & $\rationumaxAGBbclump$ & 0.761 $\pm$ 0.061 & 0.761 $\pm$ 0.062 & 0.736 $\pm$ 0.059 & 0.736 $\pm$ 0.074 \\
                                                     \hline
\label{table:ratio_AGBb_clump_location_0_6_0_9M}
\end{tabular}
\end{center}
\textbf{Notes:} Models have been obtained by individually changing the parameters of the reference model. These changes are indicated in the first column.
\end{table*}

\begin{table*}[]
\begin{center}
\caption{Ratio between the AGBb location and the clump location for models with $M \in [0.9, 1.2]M_{\odot}$}
\begin{tabular}{llllll}
\hline
\hline
$M$ ($M_{\odot}$) & & & $[0.9,1.2]$ & & \\
\hline
[Fe/H] (dex) & &  $[-1.0,-0.5]$ & $[-0.5,-0.25]$ & \ $[-0.25,0.0]$ & \ \ \ $ [0.0,0.25]$ \\
\hline
Reference model                                      & $\ratioTeffAGBbclump$  & 0.995 $\pm$ 0.004 & 0.994 $\pm$ 0.005 & 0.993 $\pm$ 0.006 & 0.993 $\pm$ 0.007 \\
                                                     & $\rationumaxAGBbclump$ & 0.804 $\pm$ 0.053 & 0.782 $\pm$ 0.061 & 0.782 $\pm$ 0.064 & 0.782 $\pm$ 0.057 \\
                                                     \hline
$\alphaMLT = 1.92 \rightarrow 1.62$                  & $\ratioTeffAGBbclump$  & 0.995 $\pm$ 0.007 & 0.993 $\pm$ 0.006 & 0.994 $\pm$ 0.007 & 0.993 $\pm$ 0.006 \\
                                                     & $\rationumaxAGBbclump$ & 0.791 $\pm$ 0.054 & 0.768 $\pm$ 0.052 & 0.768 $\pm$ 0.049 & 0.745 $\pm$ 0.049 \\
                                                     \hline
$\eta_{R} = 0.3 \rightarrow 0.1$                     & $\ratioTeffAGBbclump$  & 0.995 $\pm$ 0.005 & 0.994 $\pm$ 0.005 & 0.993 $\pm$ 0.006 & 0.993 $\pm$ 0.006 \\
                                                     & $\rationumaxAGBbclump$ & 0.804 $\pm$ 0.055 & 0.804 $\pm$ 0.056 & 0.760 $\pm$ 0.063 & 0.760 $\pm$ 0.061 \\
                                                     \hline
$\alphaovHe = 0 \rightarrow 0.5$                     & $\ratioTeffAGBbclump$  & 0.990 $\pm$ 0.006 & 0.988 $\pm$ 0.006 & 0.987 $\pm$ 0.008 & 0.986 $\pm$ 0.008 \\
                                                     & $\rationumaxAGBbclump$ & 0.589 $\pm$ 0.052 & 0.567 $\pm$ 0.039 & 0.530 $\pm$ 0.049 & 0.530 $\pm$ 0.064 \\
                                                     \hline
$\alphaovHe = 0 \rightarrow 1.0$                     & $\ratioTeffAGBbclump$  & 0.987 $\pm$ 0.006 & 0.984 $\pm$ 0.007 & 0.984 $\pm$ 0.007 & 0.982 $\pm$ 0.008 \\
                                                     & $\rationumaxAGBbclump$ & 0.445 $\pm$ 0.086 & 0.424 $\pm$ 0.082 & 0.403 $\pm$ 0.081 & 0.392 $\pm$ 0.080 \\
                                                     \hline
$\alphaovH = 0.2 \rightarrow 0.4$                    & $\ratioTeffAGBbclump$  & 0.996 $\pm$ 0.005 & 0.994 $\pm$ 0.005 & 0.994 $\pm$ 0.006 & 0.993 $\pm$ 0.004 \\
                                                     & $\rationumaxAGBbclump$ & 0.805 $\pm$ 0.086 & 0.783 $\pm$ 0.075 & 0.781 $\pm$ 0.057 & 0.781 $\pm$ 0.061 \\
                                                     \hline
$\alphaovH = 0.2 \rightarrow 0.6$                    & $\ratioTeffAGBbclump$  & 0.995 $\pm$ 0.004 & 0.995 $\pm$ 0.005 & 0.993 $\pm$ 0.006 & 0.994 $\pm$ 0.007 \\
                                                     & $\rationumaxAGBbclump$ & 0.764 $\pm$ 0.093 & 0.762 $\pm$ 0.113 & 0.762 $\pm$ 0.089 & 0.762 $\pm$ 0.087 \\
                                                     \hline
$\alpha_{\mathrm{pen\ conv,He}} = 0 \rightarrow 0.5$ & $\ratioTeffAGBbclump$  & 0.990 $\pm$ 0.006 & 0.988 $\pm$ 0.006 & 0.987 $\pm$ 0.008 & 0.986 $\pm$ 0.008 \\
                                                     & $\rationumaxAGBbclump$ & 0.589 $\pm$ 0.052 & 0.567 $\pm$ 0.039 & 0.530 $\pm$ 0.049 & 0.530 $\pm$ 0.064 \\
                                                     \hline
$\alphath = 0 \rightarrow 2$                         & $\ratioTeffAGBbclump$  & 0.996 $\pm$ 0.004 & 0.995 $\pm$ 0.005 & 0.994 $\pm$ 0.006 & 0.994 $\pm$ 0.007 \\
                                                     & $\rationumaxAGBbclump$ & 0.804 $\pm$ 0.065 & 0.804 $\pm$ 0.066 & 0.804 $\pm$ 0.063 & 0.782 $\pm$ 0.065 \\
                                                     \hline
$\alphaundersh = 0 \rightarrow 0.3$                  & $\ratioTeffAGBbclump$  & 0.996 $\pm$ 0.005 & 0.994 $\pm$ 0.005 & 0.993 $\pm$ 0.006 & 0.993 $\pm$ 0.006 \\
                                                     & $\rationumaxAGBbclump$ & 0.804 $\pm$ 0.059 & 0.782 $\pm$ 0.045 & 0.782 $\pm$ 0.044 & 0.782 $\pm$ 0.064 \\
                                                     \hline
$Y_{0} = 0.253 \rightarrow 0.303$                    & $\ratioTeffAGBbclump$  & 0.996 $\pm$ 0.004 & 0.995 $\pm$ 0.004 & 0.994 $\pm$ 0.006 & 0.993 $\pm$ 0.006 \\
                                                     & $\rationumaxAGBbclump$ & 0.786 $\pm$ 0.069 & 0.786 $\pm$ 0.065 & 0.761 $\pm$ 0.068 & 0.739 $\pm$ 0.068 \\
                                                     \hline
\label{table:ratio_AGBb_clump_location_0_9_1_2M}
\end{tabular}
\end{center}
\textbf{Notes:} Models have been obtained by individually changing the parameters of the reference model. These changes are indicated in the first column.
\end{table*}

\begin{table*}[]
\begin{center}
\caption{Ratio between the AGBb location and the clump location for models with $M \in [1.2, 1.5]M_{\odot}$}
\begin{tabular}{llllll}
\hline
\hline
$M$ ($M_{\odot}$) & & & $[1.2,1.5]$ & & \\
\hline
[Fe/H] (dex) & &  $[-1.0,-0.5]$ & $[-0.5,-0.25]$ & \ $[-0.25,0.0]$ & \ \ \ $ [0.0,0.25]$ \\
\hline
Reference model                                      & $\ratioTeffAGBbclump$  & 0.995 $\pm$ 0.005 & 0.995 $\pm$ 0.004 & 0.994 $\pm$ 0.006 & 0.995 $\pm$ 0.006 \\
                                                     & $\rationumaxAGBbclump$ & 0.821 $\pm$ 0.067 & 0.799 $\pm$ 0.091 & 0.779 $\pm$ 0.057 & 0.800 $\pm$ 0.055 \\
                                                     \hline
$\alphaMLT = 1.92 \rightarrow 1.62$                  & $\ratioTeffAGBbclump$  & 0.997 $\pm$ 0.006 & 0.994 $\pm$ 0.006 & 0.995 $\pm$ 0.007 & 0.994 $\pm$ 0.007 \\
                                                     & $\rationumaxAGBbclump$ & 0.846 $\pm$ 0.088 & 0.824 $\pm$ 0.054 & 0.801 $\pm$ 0.049 & 0.801 $\pm$ 0.049 \\
                                                     \hline
$\eta_{R} = 0.3 \rightarrow 0.1$                     & $\ratioTeffAGBbclump$  & 0.995 $\pm$ 0.004 & 0.995 $\pm$ 0.005 & 0.994 $\pm$ 0.006 & 0.993 $\pm$ 0.006 \\
                                                     & $\rationumaxAGBbclump$ & 0.821 $\pm$ 0.086 & 0.821 $\pm$ 0.093 & 0.799 $\pm$ 0.095 & 0.777 $\pm$ 0.081 \\
                                                     \hline
$\alphaovHe = 0 \rightarrow 0.5$                     & $\ratioTeffAGBbclump$  & 0.990 $\pm$ 0.006 & 0.989 $\pm$ 0.004 & 0.987 $\pm$ 0.008 & 0.986 $\pm$ 0.008 \\
                                                     & $\rationumaxAGBbclump$ & 0.614 $\pm$ 0.083 & 0.598 $\pm$ 0.079 & 0.622 $\pm$ 0.053 & 0.567 $\pm$ 0.041 \\
                                                     \hline
$\alphaovHe = 0 \rightarrow 1.0$                     & $\ratioTeffAGBbclump$  & 0.987 $\pm$ 0.006 & 0.986 $\pm$ 0.005 & 0.985 $\pm$ 0.008 & 0.983 $\pm$ 0.007 \\
                                                     & $\rationumaxAGBbclump$ & 0.495 $\pm$ 0.096 & 0.461 $\pm$ 0.095 & 0.469 $\pm$ 0.061 & 0.469 $\pm$ 0.064 \\
                                                     \hline
$\alphaovH = 0.2 \rightarrow 0.4$                    & $\ratioTeffAGBbclump$  & 0.995 $\pm$ 0.005 & 0.994 $\pm$ 0.005 & 0.994 $\pm$ 0.004 & 0.993 $\pm$ 0.006 \\
                                                     & $\rationumaxAGBbclump$ & 0.802 $\pm$ 0.105 & 0.762 $\pm$ 0.080 & 0.763 $\pm$ 0.105 & 0.782 $\pm$ 0.070 \\
                                                     \hline
$\alphaovH = 0.2 \rightarrow 0.6$                    & $\ratioTeffAGBbclump$  & 0.995 $\pm$ 0.004 & 0.994 $\pm$ 0.004 & 0.994 $\pm$ 0.005 & 0.993 $\pm$ 0.007 \\
                                                     & $\rationumaxAGBbclump$ & 0.782 $\pm$ 0.106 & 0.763 $\pm$ 0.085 & 0.763 $\pm$ 0.092 & 0.762 $\pm$ 0.113 \\
                                                     \hline
$\alpha_{\mathrm{pen\ conv,He}} = 0 \rightarrow 0.5$ & $\ratioTeffAGBbclump$  & 0.990 $\pm$ 0.006 & 0.989 $\pm$ 0.004 & 0.987 $\pm$ 0.008 & 0.986 $\pm$ 0.008 \\
                                                     & $\rationumaxAGBbclump$ & 0.614 $\pm$ 0.083 & 0.598 $\pm$ 0.079 & 0.622 $\pm$ 0.053 & 0.567 $\pm$ 0.041 \\
                                                     \hline
$\alphath = 0 \rightarrow 2$                         & $\ratioTeffAGBbclump$  & 0.996 $\pm$ 0.005 & 0.995 $\pm$ 0.004 & 0.995 $\pm$ 0.005 & 0.995 $\pm$ 0.007 \\
                                                     & $\rationumaxAGBbclump$ & 0.821 $\pm$ 0.065 & 0.799 $\pm$ 0.085 & 0.821 $\pm$ 0.058 & 0.821 $\pm$ 0.063 \\
                                                     \hline
$\alphaundersh = 0 \rightarrow 0.3$                  & $\ratioTeffAGBbclump$  & 0.995 $\pm$ 0.005 & 0.995 $\pm$ 0.004 & 0.994 $\pm$ 0.005 & 0.995 $\pm$ 0.007 \\
                                                     & $\rationumaxAGBbclump$ & 0.821 $\pm$ 0.087 & 0.821 $\pm$ 0.078 & 0.821 $\pm$ 0.047 & 0.799 $\pm$ 0.043 \\
                                                     \hline
$Y_{0} = 0.253 \rightarrow 0.303$                    & $\ratioTeffAGBbclump$  & 0.995 $\pm$ 0.005 & 0.995 $\pm$ 0.005 & 0.994 $\pm$ 0.006 & 0.994 $\pm$ 0.005 \\
                                                     & $\rationumaxAGBbclump$ & 0.823 $\pm$ 0.135 & 0.800 $\pm$ 0.144 & 0.777 $\pm$ 0.099 & 0.777 $\pm$ 0.108 \\
                                                     \hline
\label{table:ratio_AGBb_clump_location_1_2_1_5M}
\end{tabular}
\end{center}
\textbf{Notes:} Models have been obtained by individually changing the parameters of the reference model. These changes are indicated in the first column.
\end{table*}

\begin{table*}[]
\begin{center}
\caption{Ratio between the AGBb location and the clump location for models with $M \in [1.5, 2.5]M_{\odot}$}
\begin{tabular}{llllll}
\hline
\hline
$M$ ($M_{\odot}$) & & & $[1.5,2.5]$ & & \\
\hline
[Fe/H] (dex) & &  $[-1.0,-0.5]$ & $[-0.5,-0.25]$ & \ $[-0.25,0.0]$ & \ \ \ $ [0.0,0.25]$ \\
\hline
Reference model                                      & $\ratioTeffAGBbclump$  & 0.996 $\pm$ 0.004 & 0.995 $\pm$ 0.004 & 0.995 $\pm$ 0.006 & 0.994 $\pm$ 0.007 \\
                                                     & $\rationumaxAGBbclump$ & 0.841 $\pm$ 0.109 & 0.823 $\pm$ 0.083 & 0.803 $\pm$ 0.075 & 0.812 $\pm$ 0.099 \\
                                                     \hline
$\alphaMLT = 1.92 \rightarrow 1.62$                  & $\ratioTeffAGBbclump$  & 0.996 $\pm$ 0.005 & 0.995 $\pm$ 0.005 & 0.994 $\pm$ 0.006 & 0.994 $\pm$ 0.008 \\
                                                     & $\rationumaxAGBbclump$ & 0.837 $\pm$ 0.105 & 0.817 $\pm$ 0.099 & 0.821 $\pm$ 0.099 & 0.798 $\pm$ 0.094 \\
                                                     \hline
$\eta_{R} = 0.3 \rightarrow 0.1$                     & $\ratioTeffAGBbclump$  & 0.996 $\pm$ 0.004 & 0.995 $\pm$ 0.003 & 0.995 $\pm$ 0.005 & 0.994 $\pm$ 0.007 \\
                                                     & $\rationumaxAGBbclump$ & 0.841 $\pm$ 0.090 & 0.803 $\pm$ 0.092 & 0.803 $\pm$ 0.085 & 0.812 $\pm$ 0.093 \\
                                                     \hline
$\alphaovHe = 0 \rightarrow 0.5$                     & $\ratioTeffAGBbclump$  & 0.990 $\pm$ 0.005 & 0.991 $\pm$ 0.004 & 0.990 $\pm$ 0.006 & 0.988 $\pm$ 0.007 \\
                                                     & $\rationumaxAGBbclump$ & 0.695 $\pm$ 0.110 & 0.646 $\pm$ 0.075 & 0.646 $\pm$ 0.070 & 0.632 $\pm$ 0.076 \\
                                                     \hline
$\alphaovHe = 0 \rightarrow 1.0$                     & $\ratioTeffAGBbclump$  & 0.989 $\pm$ 0.005 & 0.988 $\pm$ 0.004 & 0.987 $\pm$ 0.006 & 0.985 $\pm$ 0.007 \\
                                                     & $\rationumaxAGBbclump$ & 0.641 $\pm$ 0.128 & 0.592 $\pm$ 0.080 & 0.560 $\pm$ 0.074 & 0.547 $\pm$ 0.091 \\
                                                     \hline
$\alphaovH = 0.2 \rightarrow 0.4$                    & $\ratioTeffAGBbclump$  & 0.997 $\pm$ 0.005 & 0.996 $\pm$ 0.004 & 0.996 $\pm$ 0.005 & 0.995 $\pm$ 0.006 \\
                                                     & $\rationumaxAGBbclump$ & 0.890 $\pm$ 0.159 & 0.823 $\pm$ 0.105 & 0.807 $\pm$ 0.101 & 0.807 $\pm$ 0.101 \\
                                                     \hline
$\alphaovH = 0.2 \rightarrow 0.6$                    & $\ratioTeffAGBbclump$  & 0.999 $\pm$ 0.007 & 0.998 $\pm$ 0.006 & 0.996 $\pm$ 0.006 & 0.996 $\pm$ 0.006 \\
                                                     & $\rationumaxAGBbclump$ & 0.865 $\pm$ 0.221 & 0.875 $\pm$ 0.178 & 0.844 $\pm$ 0.131 & 0.801 $\pm$ 0.118 \\
                                                     \hline
$\alpha_{\mathrm{pen\ conv,He}} = 0 \rightarrow 0.5$ & $\ratioTeffAGBbclump$  & 0.990 $\pm$ 0.005 & 0.991 $\pm$ 0.004 & 0.990 $\pm$ 0.006 & 0.988 $\pm$ 0.007 \\
                                                     & $\rationumaxAGBbclump$ & 0.695 $\pm$ 0.110 & 0.646 $\pm$ 0.075 & 0.646 $\pm$ 0.070  & 0.632 $\pm$ 0.076 \\
                                                     \hline
$\alphath = 0 \rightarrow 2$                         & $\ratioTeffAGBbclump$  & 0.997 $\pm$ 0.004 & 0.995 $\pm$ 0.003 & 0.995 $\pm$ 0.005 & 0.994 $\pm$ 0.008 \\
                                                     & $\rationumaxAGBbclump$ & 0.841 $\pm$ 0.127 & 0.823 $\pm$ 0.101 & 0.805 $\pm$ 0.086 & 0.813 $\pm$ 0.089 \\
                                                     \hline
$\alphaundersh = 0 \rightarrow 0.3$                  & $\ratioTeffAGBbclump$  & 0.996 $\pm$ 0.005 & 0.995 $\pm$ 0.004 & 0.995 $\pm$ 0.005 & 0.994 $\pm$ 0.007 \\
                                                     & $\rationumaxAGBbclump$ & 0.841 $\pm$ 0.101 & 0.842 $\pm$ 0.085 & 0.803 $\pm$ 0.095 & 0.812 $\pm$ 0.087 \\
                                                     \hline
$Y_{0} = 0.253 \rightarrow 0.303$                    & $\ratioTeffAGBbclump$  & 0.998 $\pm$ 0.005 & 0.996 $\pm$ 0.005 & 0.995 $\pm$ 0.005 & 0.995 $\pm$ 0.006 \\
                                                     & $\rationumaxAGBbclump$ & 0.859 $\pm$ 0.197 & 0.840 $\pm$ 0.124 & 0.820 $\pm$ 0.103 & 0.810 $\pm$ 0.081 \\
                                                     \hline
$\Omega_{\mathrm{ZAMS}}/\Omegacrit = 0 \rightarrow 0.3$                       & $\ratioTeffAGBbclump$  & 0.999 $\pm$ 0.004 & 0.995 $\pm$ 0.005 & 0.995 $\pm$ 0.005 & 0.994 $\pm$ 0.006 \\
                                                 & $\rationumaxAGBbclump$ & 0.820 $\pm$ 0.200    & 0.754 $\pm$ 0.150  & 0.783 $\pm$ 0.174 & 0.777 $\pm$ 0.118 \\
\hline
test                      & $\ratioTeffAGBbclump$  & 0.985 $\pm$ 0.007 & 0.985 $\pm$ 0.012 & 0.984 $\pm$ 0.010  & 0.985 $\pm$ 0.007 \\
                                                & $\rationumaxAGBbclump$  & 0.733 $\pm$ 0.275 & 0.508 $\pm$ 0.136 & 0.464 $\pm$ 0.183 & 0.474 $\pm$ 0.096 \\
\hline
\label{table:ratio_AGBb_clump_location_1_5_2_5M}
\end{tabular}
\end{center}
\textbf{Notes:} Models have been obtained by individually changing the parameters of the reference model. These changes are indicated in the first column. The model labelled `test' is obtained with the following changes: $\Omega_{\mathrm{ZAMS}}/\Omegacrit = 0 \rightarrow 0.3$, $\alphaovH = 0.2 \rightarrow 0$, $\alphaovHe = 0 \rightarrow 1.0$, and $\alphaMLT = 1.92 \rightarrow 1.62$.
\end{table*}

\end{appendix}

\end{document}